\documentclass[fleqn,10pt]{wlscirep}
\usepackage[utf8]{inputenc}
\usepackage[T1]{fontenc}
\usepackage{booktabs}
\usepackage{array}
\usepackage{multirow}
\usepackage{url}
\usepackage{graphicx}
\usepackage{comment}
\usepackage{xr} 
\usepackage[colorlinks=true, allcolors=blue]{hyperref}
\usepackage{subcaption}
\usepackage{orcidlink}
\usepackage{amsmath,amssymb}
\usepackage{booktabs}
\usepackage{hyperref}
\usepackage{textcomp}

\usepackage[table]{xcolor}
\definecolor{lightgreen}{RGB}{220,240,220}
\definecolor{lightred}{RGB}{245,220,220}
\usepackage{tabularx}
\usepackage{array}
\usepackage{makecell} 

\makeatletter
\newcommand*{\addFileDependency}[1]{
  \typeout{(#1)}
  \@addtofilelist{#1}
  \IfFileExists{#1}{}{\typeout{No file #1.}}
}
\makeatother
\newcommand*{\myexternaldocument}[1]{%
    \externaldocument{#1}%
    \addFileDependency{#1.tex}%
    \addFileDependency{#1.aux}%
}
\myexternaldocument{main}

\usepackage[resetlabels]{multibib}

\title{Bringing the economics of biodiversity into policy and decision-making: A target and cost-based approach to pricing biodiversity}

\author[1,2,*]{Ben Groom\orcidlink{0000-0003-0729-143X}}
\author[3]{Joseph Lowe}
\author[4]{Sophus zu Ermgassen\orcidlink{0000-0001-6044-3389}}
\author[4]{E.\,J. Milner-Gulland\orcidlink{0000-0003-0324-2710}}
\author[4]{Thomas Atkins}
\author[1]{Ben Balmford\orcidlink{0000-0002-4102-5250}}
\author[1]{Amy Binner\orcidlink{0000-0003-3235-0585}}
\author[4]{Amber Butler}
\author[1]{Brett Day\orcidlink{0000-0001-7519-5672}}
\author[4]{Natalie Duffus\orcidlink{0000-0001-7126-4909}}
\author[5]{Rosie Hails\orcidlink{0000-0002-6975-1318}}
\author[2]{Hannah Maier-Peveling}
\author[1]{Mattia Mancini\orcidlink{0000-0003-0025-2462}}
\author[1]{Sarah Meier\orcidlink{0000-0001-6108-0735}}
\author[4]{Hannah Nicholas}
\author[1]{Daniele Rinaldo\orcidlink{0000-0002-5312-5697}}
\author[6]{Robin Smale}
\author[7]{Pat Snowdon}
\author[2]{Frank Venmans\orcidlink{0000-0002-4264-6606}}
\author[1,*]{Ian J.\ Bateman\orcidlink{0000-0002-2791-6137}}

\affil[1]{LEEP Institute, Department of Economics, University of Exeter, Rennes Drive, Exeter, EX4 4RJ, UK}
\affil[2]{Grantham Research Institute on Climate Change and the Environment, London School of Economics, Houghton Street, London, WC2A 2AE, UK}
\affil[3]{OPEN Fellow, Department of Biology, University of Oxford, UK}
\affil[4]{Department of Biology, University of Oxford, UK}
\affil[5]{National Trust, UK}
\affil[6]{Freelance Economist, UK}
\affil[7]{Scottish Forestry, Scottish Government, UK}

\affil[*]{Corresponding authors: b.d.groom@exeter.ac.uk, i.bateman@exeter.ac.uk}

\begin{abstract}
Given ongoing, human-induced, loss of wild species we propose the Target and Cost Analysis (TCA) approach as a means of incorporating biodiversity within government appraisals of public spending. Influenced by how carbon is priced in countries around the world, the resulting biodiversity `shadow' price reflects the marginal cost of meeting government targets while avoiding disagreements on the use of willingness to pay measures to value biodiversity. Examples of how to operationalize TCA are developed at different scales and for alternative biodiversity metrics, including extinction risk for Europe and species richness in the UK. Pricing biodiversity according to agreed targets allows trade-offs with other wellbeing-enhancing uses of public funds to be sensibly undertaken without jeopardizing those targets, and is compatible with international guidelines on Cost Benefit Analysis.
\end{abstract}

\begin{document}
\flushbottom
\maketitle

\noindent Respecting planetary boundaries for biodiversity defines a key dimension of sustainability \cite{Dasgupta2021Biodiversity,Steffen2015PlanetaryBoundaries}. Recognising this, numerous international agreements, stretching from the Earth Summit of 1992 to the recent Kunming-Montreal Agreement of the Convention on Biological Diversity of 2023\cite{UN2022KunmingMontreal}, have sought to `bend the curve' of biodiversity loss \cite{Leclere2020BendingCurve}. However, to date such agreements have been insufficient, with losses still ongoing globally\cite{Keck2025HumanImpact}. Given this, targets 8-18 of the Kunming-Montreal Agreement emphasise the need to extend consideration of biodiversity beyond international treaties and the traditional confines of conservation policy. Rather these targets seek to bring biodiversity into general decision-making across areas (including agriculture, forestry, climate change mitigation, infrastructure and planning) that both directly and indirectly affect wild species. However, this crucial mainstreaming of biodiversity across all areas of policy is hampered by the fact that the value of biodiversity and the costs of its loss are routinely omitted from the everyday economic Cost-Benefit Analysis (CBA) assessments which underpin much government decision-making internationally \cite{HMTreasury2022GreenBook,EC2014CBAGuide, Robinson2019BCAGuidelines}. Effectively biodiversity becomes invisible at the point of decision such that insufficient resources are allocated to combat its decline\cite{ipbes_global_2019}, while at the extreme biodiversity loss is effectively subsidised\cite{Dasgupta2021Biodiversity}. A coherent and readily generalisable approach is needed which delivers biodiversity targets while allowing due weight to the other values generated by public expenditure. To this end we build on existing international approaches to the incorporation of a carbon `shadow' price \cite{Hof2017NDCCosts} within decision-making to propose the Target and Cost Analysis (TCA) approach, in which a price for biodiversity is defined to reflect the marginal cost of delivering societally defined targets for biodiversity restoration. The target-compatible `shadow price' emerging from the TCA approach provides a basis upon which to consistently evaluate biodiversity-economy-wellbeing trade-offs across government, and a benchmark for private sector action.

Just as reliance on the shadow price for carbon reflects the overwhelming complexity of estimating a robust value for avoiding the myriad impacts of climate change, so the shadow price provided by the TCA approach provides a pragmatic response to the challenge of estimating a defensible value for biodiversity. Economic CBA assessments examine the marginal (i.e. per unit) net benefit of a change in resource use. However, the lack of knowledge regarding the quantified connection between the biodiversity of a system and the complex ecosystem services and processes it supports precludes such marginal valuations \cite{Dasgupta2021Biodiversity,IPBES2022ValuesNature,Jochum2020BEFExperiments}. Similarly, survey or experimental attempts to estimate individuals' willingness to pay (WTP) values yield responses which are insensitive to marginal changes and are instead influenced by psychological heuristics \cite{Horisch2024BiodiversityWTP}. Tail risks (low risk, but potentially catastrophic events) also undermine benefits estimation more generally \cite{Weitzman2009CatastrophicClimate}, while non-economic critiques challenge the Utilitarian basis of value \cite{IPBES2022ValuesNature}. Such challenges can act as a barrier to agreement on policy. The TCA approach avoids these `demand-side' methodological and ethical criticisms of economic valuation. Rather than reflecting often unreliable and incomplete estimates of WTP, the TCA leads to a shadow price for biodiversity that simply reflects the marginal cost of meeting a societally agreed target for biodiversity. The target could be chosen in consultation with experts in the field and other stakeholders. A prudent quantitative constraint on minimum levels of restoration would embody concerns about the aforementioned uncertainties associated with biodiversity and a combination of consequential and deontological (rights and duties) reasoning. The emerging shadow price can then be used to guide public investment and regulatory appraisal and across all government departments, correcting for otherwise unaccounted-for harms or benefits to biodiversity in a way that reflects agreed national objectives for biodiversity, not solely WTP. A central accounting or shadow price for biodiversity would mainstream biodiversity in decision-making in the same way that the value of time saved or the value of an avoided fatality are already, rather than biodiversity being systematically overlooked.

Precisely because of the difficulties outlined above, the UK Treasury is considering the TCA proposals put forward here as a means of incorporating biodiversity within the so-called Green Book \cite{HMTreasury2022GreenBook}, the UK official guidance on CBA for public spending by government departments. The TCA approach proposed in this paper is seen as a complement to the Green Book's natural capital framework for appraising public investment and the Enabling Natural Capital Approach (ENCA) guidelines of the UK Department of Environment, Food and Rural Affairs \cite{DEFRA2023ENCA} (members of which are also co-authors of this paper). TCA also provides a conduit for the UK Government's 2021 Environment Act \cite{UKGov2021EnvironmentAct} whether implemented through targets for improving species abundance, lowering extinction risk or increasing habitable area. The approach can be seen as part of the UK Government's broad response to the Dasgupta Review \cite{HMTreasury2021DasguptaResponse} and has general applicability as a means of incorporating biodiversity within economic and policy decision-making across all government departments, in other sectors, but also worldwide.

The principles of the TCA approach to pricing biodiversity are outlined in the following section and its challenges discussed. Examples are provided to illustrate how the approach could work in practice across different administrative scales and target metrics. The first example appraises the entire European Economic Area using extinction risk as a metric, while a second example considers imagined national level targets for species richness using a spatially optimized analysis. We then show how the approach can be applied to evaluate a local development in a typical Cost Benefit Analysis. Together these examples demonstrate \textit{proof of concept} for deployment of TCA in the appraisal of public investments and regulatory changes to augment CBA and Natural Capital Accounting guidelines \cite{DEFRA2023ENCA}.  Policy ready numbers require further engagement with policy makers and alignment on metrics and methods, which is ongoing in the UK.

\section*{Principles of the Target and Cost Analysis (TCA) approach}

The TCA approach for pricing biodiversity shares key principles with current methods for pricing carbon \cite{DESNZ2024EnergyGHGValuation,WorldBank2024CarbonShadowPrice}, albeit with some important differences in practice: e.g. the metric for climate policy is more obvious (carbon emissions or temperature change) than for biodiversity. In the carbon context, economic cost-benefit assessments of climate policy would ideally be guided by balancing the costs of avoiding the emission of an additional (or `marginal') tonne of carbon with its benefits, namely the present value of the avoided future damages which would have arisen if that tonne of carbon had been emitted (the latter known as the Social Cost of Carbon; SCC) \cite{Aldy2021SCC}. Unfortunately (just like the theoretical true value of biodiversity), the SCC is very difficult to estimate robustly, being highly sensitive to uncertainties regarding the level of future climate damages\cite{Pindyck2013JEL} and global policy response\cite{IPCC2023AR6SynthesisSections}, the social discount rate \cite{Drupp2018Discounting}, climate tipping points \cite{Dietz2021TippingPoints} and the measurement of ecosystem damages \cite{DruppHansel2021RelativePrices}. In response, the EU and countries such as China, India and the UK \cite{UKGov2019ClimateActAmendment} have set policy targets to achieve specified reductions in emissions by specified dates, eventually reaching net zero. The cost of meeting targets is then established by examining alternative emissions reduction technologies such as energy efficiency measures, renewable energy, sequestration via nature-based solutions, etc. Each of these has a different per tonne cost for abating carbon, known as its Marginal Abatement Cost (MAC). By ordering the various technologies available for abating carbon in terms of ascending marginal cost per tonne of removal we generate a MAC curve as illustrated in Fig. 1.

\begin{figure}[h]
\centering
\includegraphics[width=0.7\linewidth]{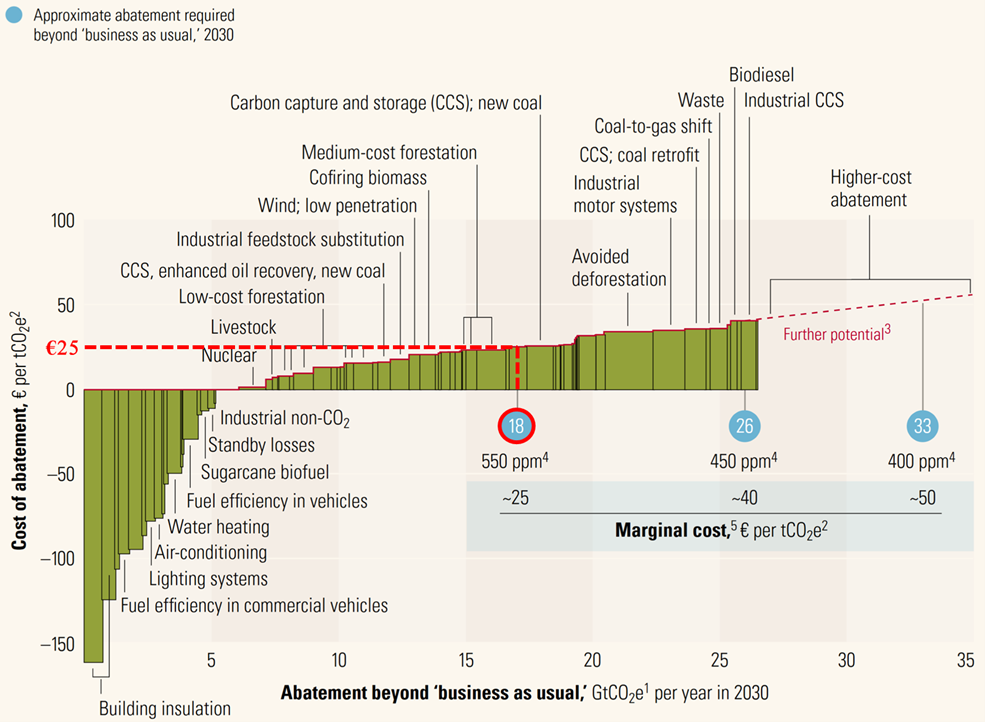}
\caption{\textbf{Global Marginal Abatement Cost (MAC) curve for greenhouse gas abatement measures.} Note, negative values appear due to the positive financial returns associated with some efficiency measures, such as insulation and fuel efficiency. This MAC curve largely ignores the transaction costs of implementation technologies. Note: $^1$ GtCO$_2$e = gigaton of carbon dioxide equivalent; The abatement measures shown are for investments beyond “business as usual” based on emissions growth driven mainly by increasing demand for energy and transport around the world and by tropical deforestation. $^2$ tCO$_2$e = ton of carbon dioxide equivalent. $^3$ Measures costing more than €40 a ton were not the focus of this study. $^4$ Atmospheric concentrations of all greenhouse gases recalculated into CO$_2$ equivalents; ppm = parts per million. $^5$ Marginal cost of avoiding emissions of 1 ton of CO$_2$ equivalents in each abatement demand scenario. Source: Adapted from \cite{Enkvist2007CostCurve}}
\label{fig:1}

\end{figure}

The appropriate carbon price in this framework reflects the MAC of reducing emissions to attain the target level. The vertical broken line in Fig 1 shows the level of emissions reduction consistent with an annual abatement that delivers the net zero commitment by its specified date (e.g. in the case of the UK, by 2050), while the horizontal broken line indicates the corresponding shadow price compatible with such a target level of abatement. Clearly more ambitious targets imply a higher carbon price. By employing carbon abatement technologies, such as energy efficiency improvements, costing less than or equal to the shadow price, a government would ensure least cost adoption of carbon reduction technologies to reach the net zero target. Similarly, policies and investments that increase carbon emissions can be costed using the shadow price since they raise the costs of achieving the target by this amount.

The MAC approach avoids the difficulties of robustly estimating the Social Cost of Carbon, thereby allowing decision makers to bring greenhouse gas emissions into policy making in a manner consistent with attaining the net zero target. The TCA framework similarly avoids the intractable challenge of estimating the true value of biodiversity by introducing a cost commensurate with attaining an agreed biodiversity target.

\section{The TCA approach and the Marginal Biodiversity Recovery Cost (MBRC) curve}

For the TCA approach to be operationalised for biodiversity a legislatively binding target is required (akin to the net-zero policy for greenhouse gases), together with a clear overarching strategy for delivering that target. Internationally, all signatories to the Kunming-Montreal Agreement have agreed to halt and reverse the loss of biodiversity by 2030 and reduce extinction risk over similar horizons; biodiversity targets are therefore globally accepted. At national level, taking the case of England as an example, candidate targets for biodiversity stem from commitments made in the 25 Year Environment Plan \cite{DEFRA2018EnvironmentPlan} and the 2021 Environment Act \cite{UKGov2021EnvironmentAct}. One such target, for example, requires that the decline in the abundance (populations) of species within England be halted by 2030 and populations increased by at least 10\% by 2042. There is also a statutory target for extinction risk for England\cite{UKGovStatutorySpeciesTargets}.

Any practical target must be framed in terms of a clearly defined metric, against which the consequences of different interventions on biodiversity can then be assessed. Unlike carbon, biodiversity does not easily lend itself to being measured by a single metric. Species richness \cite{baumgartner_measuring_2006}, extinction risk \cite{Mace2008IUCNExtinctionRisk}, species recovery \cite{Akcakaya2018IUCNGreenList}, population and habitat intactness \cite{jetz_include_2022} and phylogenetic distinctiveness \cite{Weitzman1992OnDiversity} are just some examples of the broad range of biodiversity metrics available. Each differs in its concept of biodiversity, its policy relevance and ease of implementation (data requirements, ease of calculation, etc.). In England a key metric operationalised in public policy is the Department of Food, Agriculture and Rural Affairs (DEFRA) Statutory Biodiversity Metric, an index that captures biodiversity via measures of habitat condition, strategic significance, feasibility, and timescale of biodiversity restoration. In England national targets have also been set for habitat area, species abundance and extinction risk, in line with the recommendations of the Environment Act (2021) \cite{UKGov2021EnvironmentAct}. Each metric and target has its own merits and we do not attempt to discriminate between alternatives here. The key point is that a target and metric can be defined in ecological terms, rather than purely economic ones. In principle TCA can handle multiple targets (See p3 of SI), but we illustrate the approach using just one.

With the metric and target defined, the next step is to estimate the biodiversity equivalent of the MAC curve for carbon: the Marginal Biodiversity Recovery Cost (MBRC) curve. Fig. 2 shows how this curve could be assembled from different biodiversity-enhancing `technologies', such as changes in land use and land management, arranged in ascending order of marginal cost (just as for carbon in Fig. 1). In Fig. 2, the horizontal axis uses the agreed metric to quantify biodiversity and shows the current and target levels. For any desired biodiversity outcome there is a corresponding marginal cost (measured on the vertical axis). The point at which the target level crosses the MBRC curve determines the appropriate, target compatible shadow price of biodiversity; the price at which the target will be attained. The more stringent the biodiversity target the further we move up the MBRC curve and the higher the price. Any projects that affect biodiversity can then be assessed using this price with those projects providing biodiversity at a cost higher than the shadow price being rejected while those at or below the shadow price being accepted. Otherwise, interventions causing external costs of biodiversity loss can now be priced-in to shape the broader investment appraisal. 

\begin{figure}[t]
\centering
\includegraphics[width=0.9\linewidth]{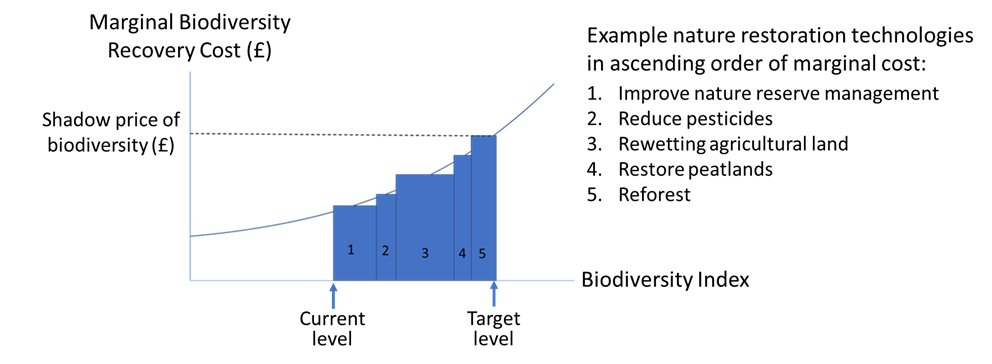}
\caption{\textbf{Marginal Biodiversity Recovery Cost (MBRC) curve.} An example of the sequence of technologies that could be used to achieve a biodiversity target.}

\label{fig:2}

\end{figure}

In principle the TCA approach is straightforward, building upon current practice in carbon pricing to provide a price that reflects overall societal priorities. The shadow price allows biodiversity to be brought into the economic analysis of any investment or policy decision without focussing solely on biodiversity valuation (WTP) exercises. The SI (p3) provides a theoretical model of the TCA in which biodiversity targets are met cost-effectively in a context with spatial interactions between land uses and habitat. This illustrates the generality of the approach to multiple targets at different spatial scales. We now consider remaining challenges and opportunities arising from implementation of the TCA approach to pricing biodiversity.

\section{Challenges and opportunities}

\subsection{Measurement and metrics}

There is a clear need to measure the effectiveness of interventions, if only to compare across options. As noted above, there are a number of biodiversity metrics which could be adopted \cite{baumgartner_measuring_2006, Akcakaya2018IUCNGreenList, Mace2008IUCNExtinctionRisk,Weitzman1992OnDiversity} each of which addresses a different aspect or definition of biodiversity. The choice of metrics should correspond to the aspects of biodiversity of concern. This is not an issue which will be solved here, rather it is a discussion which must take place, irrespective of the TCA (or any other) approach to bringing biodiversity into decision-making. In the meantime, metrics such as the UK Government's DEFRA Statutory Biodiversity Metric and international targets (e.g. Kunming-Montreal Agreement) and associated metrics related to increasing species abundance, reducing extinction risk and restoring habitat, are already in use in public policy. The choice of metric is important, but current use of metrics and targets shows that this decision should not rule out TCA. Below we detail practical examples illustrating how the TCA approach can be deployed using different metrics. Our local case study (see p24 of SI) provides an application at the project level using the statutory metric for the Biodiversity Net Gain policy.

\subsection{Fungibility and substitutability}

The geographical location of carbon emissions or sequestration does not matter in terms of their impact on climate change. This is not the case for biodiversity. Habitats and their biodiversity differ from one place to another. Changing extinction risk in one place affects a different set of species than a change in another. Substitutability, irreversibility and the irreplaceability of components of biodiversity become important. Some jurisdictions use a `mitigation hierarchy' framework which should in theory prevent unique, non-substitutable or irreplaceable biodiversity from being lost. In terms of the TCA approach, separate targets and corresponding MBRCs could be defined for given elements of biodiversity to reflect non-substitutability. For example, woodland losses might require woodland offsets, losses of at-risk species may be non-fungible and require same-species offsets, while losses of an abundant species might justify substitution with different, at-risk species \cite{Arlidge2018MitigationHierarchy}. The theoretical example in the SI (p3) captures these issues, including the importance of contiguity of habitat, within a matrix of spatial interactions, illustrating how these issues can in principle be embodied in the TCA approach.

Applying the TCA approach across a wide area, such as at a national scale, will maximise potential efficiencies and the level of biodiversity gain achievable for any given budget. In contrast, confining TCA assessments and biodiversity offset projects to smaller areas necessarily reduces the set of possible solutions and raises the costs of achieving a set target \cite{Mancini2024Offsets}. Looked at another way, the higher costs of such spatial constraints in part reflect the fact that certain aspects of biodiversity are not substitutable for one another. The higher restoration costs would then reflect the best available knowledge on (non-)substitutability between different aspects of biodiversity \cite{Mancini2024Offsets}. As spatial constraints tighten, the costs associated with a like-for-like replacement at the level and in the environs of the intervention become appropriate for investment and policy appraisal. This reduces the TCA approach to a replacement cost approach, an accepted approach in public appraisal guidelines such as ENCA\cite{DEFRA2023ENCA}. Such a \textit{localised} example is explored in the SI (p24).

\subsection{Uncertainty and sensitivity}

As is the case for the carbon MAC \cite{kesicki_marginal_2012}, there are likely to be uncertainties in the estimation of the MBRC derived from sampling, measurement errors, and model uncertainty. For instance, the precise cost of different land use changes, input substitutions, or biodiversity-enhancing technology are not known with certainty ex-ante. Such uncertainties and measurement errors need to be acknowledged and assessed for any systematic biases, particularly where the ordering of technologies affects the shadow price at the target level of biodiversity. As with any decision, decision makers ought to be presented with some information regarding the precision of estimates and sources of uncertainty. TCA has some advantages over WTP approaches, since the cost of land and restoration is independently observable and verifiable and will become better understood over time as policies are undertaken and more is learned about the relationship between land-use change and restoration. The same cannot always be said about marginal benefits of complex objects, such as the damages from climate change or biodiversity loss, whose value depends on often unknowable, labile, and ambiguous aspects of environment, ecology, and economy\cite{Pindyck2013JEL}. Where ecological modelling (rather than statistical) uncertainty is an issue, sensitivity analysis is appropriate to reflect modelling uncertainty. The literature on the species-area relationship (SAR), which we explore below, suggests that there is uncertainty concerning the functional form and parameterisation of SARs \cite{BodeMurdoch2009SARRobustness}, albeit with varying impact on policy recommendations \cite{Guilhaumon2008SARUncertainty}. Such uncertainty will affect the estimated shadow price. In the applications that follow we illustrate these sensitivities. For the European application we discuss uncertainty in the modelling of extinction risk and illustrate the impact of previous estimates of the 95\% confidence intervals for the SAR. In the UK application we explore how different biodiversity-enhancing technologies affect outcomes, reflecting sensitivity of shadow prices to the technology deployed. Each case illustrates the need for further research to reduce or better characterise parametric and model uncertainties. Alternatively, operationalising TCA will require an analysis of the cumulative uncertainty of modelling complex systems. Future applications could use emulation of Gaussian methods \cite{MingWilliamson2023LinkedDeepGP, MingWilliamsonGuillas2023DeepGPImputation} to quantify, report and reduce modelling uncertainty on shadow price estimates.

\section{Estimating the MBRC curve}

As shown in Fig 2, the key information for estimating the MBRC is the connection between the costs of the biodiversity-enhancing technology and the targeted biodiversity metric. Conservation science and economics have long been concerned with estimating cost-effective strategies for conservation\cite{Withey2012ConservationROI, Ando1998SpeciesDistributions}. Three examples illustrate how the MBRC could be estimated and target-compatible pricing undertaken at different scales: (i) Extinction risk-reduction targets for Europe; (ii) Species richness in the UK; (iii) A local UK example using the statutory DEFRA biodiversity metric in the context of Biodiversity Net Gain (BNG) (See p24 of SI). In each case the MBRC curve is calculated by first assessing the cost-effectiveness of land use changes and identifying those areas which provide the largest increase in the biodiversity metric per unit cost. In (i) and (ii) we focus on the opportunity cost of land, which typically dominates total costs \cite{Bateman2015MarketForcesSpatialTargeting}, setting restoration costs to zero because of data limitations. The MBRC is the inverse of this ratio: the marginal cost per unit change in the metric. The Methods and SI (p3) provide formal details.

\subsection{A persistence / extinction-risk MBRC for the European Economic Area (EEA)}

Signatories to the Kunming-Montreal agreement committed to reduce extinction risk by ten percent by 2030 and to zero by 2050. Indeed, targets in relation to extinction risk have long existed at regional administrative scales such as the European Union \cite{Natura2000Overview} and at country-level, e.g. the UK, where there are statutory target for extinction risk and species abundance \cite{UKGovStatutorySpeciesTargets, UKGov2023EnvironmentalImprovementPlan}. Extinction risk is an obvious context in which to illustrate proof-of-concept for estimating the MBRC.

\begin{figure}[h]
\centering
\begin{subfigure}{0.48\linewidth}
  \centering
  \includegraphics[width=\linewidth]{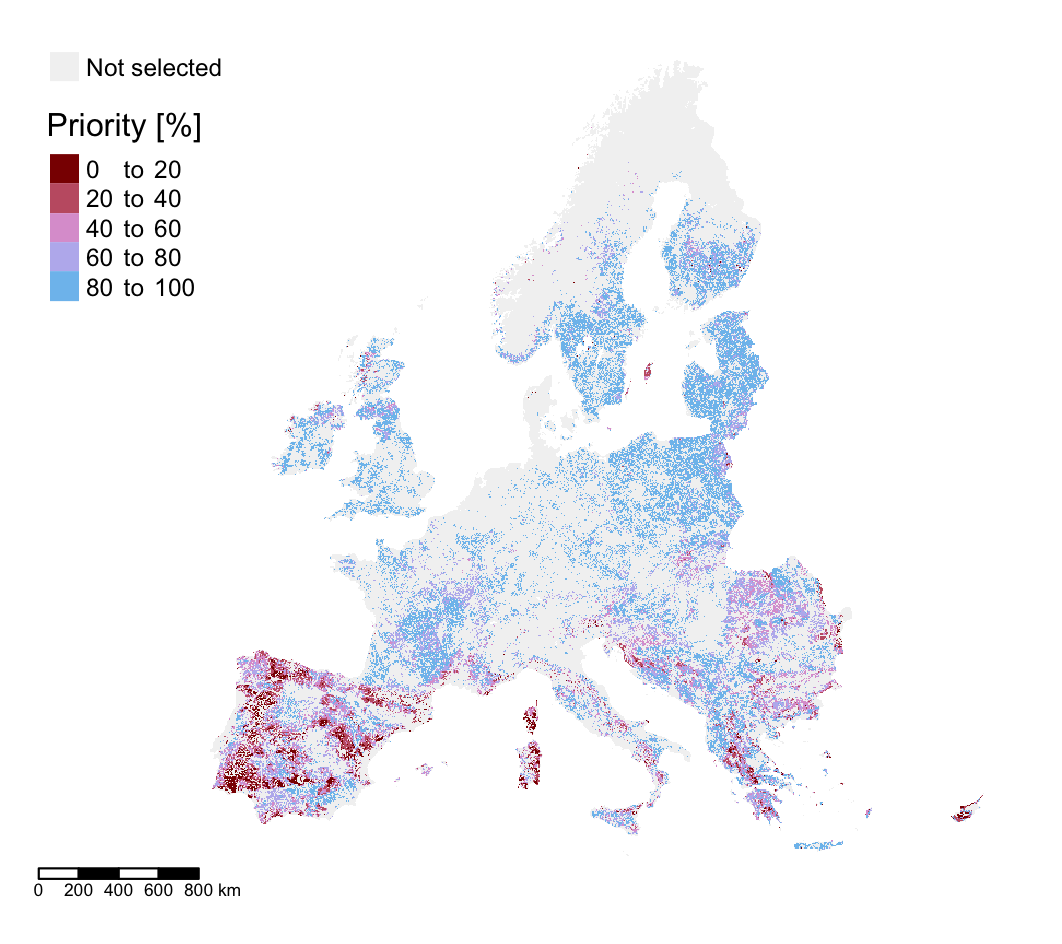}
  \caption{}
  \label{fig:3a}
\end{subfigure}
\begin{subfigure}{0.48\linewidth}
  \centering
  \includegraphics[width=\linewidth]{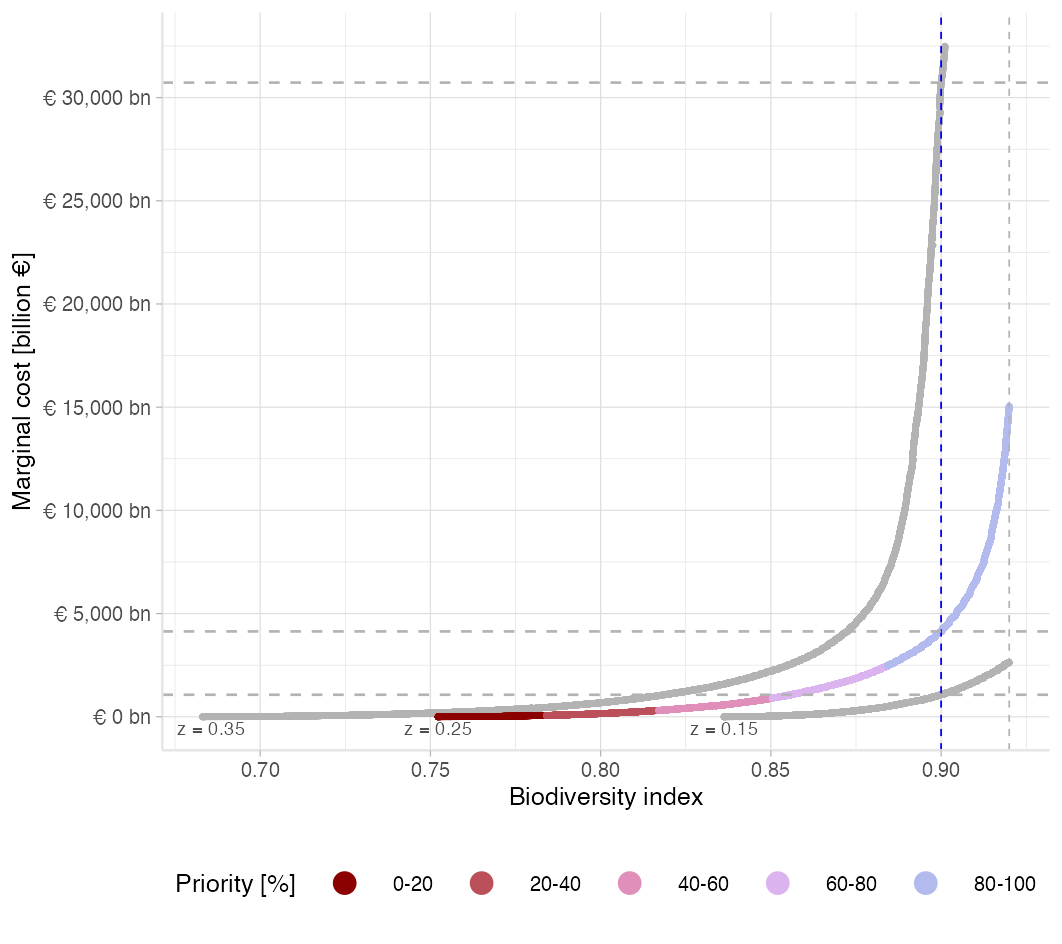}
  \caption{}
  \label{fig:3b}
\end{subfigure}
\caption{Cost effective reductions in extinction risk and the Marginal Biodiversity Recovery Cost (MBRC) curve: Fig. \ref{fig:3a}  shows the cost-effective plots that could be targeted to reduce extinction risk in the EEA zone using the relationship Persistence $p=(H/OH)^{z}$, under the assumption that $z = 0.25$ (see Equation 1 in Methods). The red areas are more cost effective than blue ones (see legend). Fig. \ref{fig:3b} shows the associated MBRC for the EEA, colour-coded in the same way as for Fig. \ref{fig:3a}, with the low cost interventions in red and higher cost interventions in blue and the X-axis measuring persistence, noting that persistence $= 1 -$ extinction risk (e.g.\ \cite{Strassburg2020RestorationPriorities}). The shadow price for biodiversity can be read off the Y-axis of Fig. \ref{fig:3b} for any target set on the X-axis, where the current level of average persistence for the EEA area is $0.75$ when $z = 0.25$. Fig. \ref{fig:3b} also shows sensitivity to the parameter $z$ with the upper and lower curves using $z = 0.35$ and $z = 0.15$, respectively the upper and lower bound of the 95\% confidence interval used in \cite{Strassburg2020RestorationPriorities}.}
\label{fig:3}

\end{figure}
A number of studies have used the relationship between the ratio of area of suitable habitat that is available to species at any point in time ($H$) and original area of Habitat ($OH$), and the risk of extinction \cite{duran_practical_2020, Strassburg2020RestorationPriorities} to identify policy levers for biodiversity conservation. Eqs. 1-6 in the Methods section explain the approach formally. Species $i$'s probability of survival, it's persistence $p_i$, is related to the proportion of its suitable habitat remaining , $H_i/OH_i$ via a power function: $p_i = (H_i/OH_i)^z$, where z reflects the sensitivity of species $i$ to the proportion of its remaining habitat \cite{Strassburg2020RestorationPriorities} (See Eq. 1 in Methods). Extinction risk is given by $1 - p_i$. 
Following \cite{Strassburg2020RestorationPriorities}, an index for the average persistence can be calculated for all species in an area of a particular granularity (e.g. 1km grid square) that is then aggregated for a defined policy jurisdiction, in this case the EEA (Eq. 2 in Methods). This relationship connects interventions that affect habitat $H$ (relative to $OH$) to a policy relevant biodiversity index: persistence and its complement extinction risk. The impact on the index can be modelled using plausible assumptions about how the presence and absence of species is affected by restoration of habitat from the current to a pristine state and what it costs to instigate that change \cite{Strassburg2020RestorationPriorities, Withey2012ConservationROI,Ando1998SpeciesDistributions}. Specifically, we assume that 1km grid squares are changed from current non-pristine habitat to pristine habitat at a cost determined by the opportunity cost of land use \cite{Spencer2024GreenInfrastructure}, i.e. the foregone current (overwhelmingly agricultural) rental value of that grid square (Eq. 5 in Methods. See also SI). Due to data limitations, restoration costs are set to zero, implying a rewilding approach. Current and pristine states of habitat are identified using maps of pre-disturbance and current habitat cover for the EEA area classified as broad (level 1) habitat types following \cite{iucn_habitats_2024} and shown in \cite{jung_assessment_2024,jung_global_2020} (See Figs. S1(a) and S1(b) on p 21 of SI) For a species to enter into the calculation of the index (Eq 2 in Methods) it must in principle be able to survive in a restored grid cell. To establish this we focus on the IUCN European Red List species and associated range maps \cite{IUCN2024EuropeanRedList,iucn_spatial_2024} and require that both habitat suitability and range maps coincide in the restored grid cell.

With cost and persistence/extinction risk measures defined, we then rank all locations (e.g. grid squares) in order of their cost-effectiveness of achieving a given change in persistence, i.e. the cost per unit reduction in the persistence index (Eq. 5 in Methods). The inverse of the cost effectiveness ratio is the MBRC for aggregate persistence, which measures the marginal cost per unit change in this biodiversity index (Eq. 6 in Methods). With a binding target level of persistence set by policy, the MBRC reflects the cost-effective sequence of restoration that leads to the target being met (similar to \cite{Strassburg2020RestorationPriorities} and \cite{Withey2012ConservationROI} for example). The value of the MBRC at the target level of persistence defines the `shadow' price. This can then be used to price changes in extinction risk in public investment appraisal and the analysis of regulatory change.

Using persistence as the biodiversity index, Fig. \ref{fig:3} illustrates the outcome of these steps in the EEA. Fig. \ref{fig:3a} is a colour coded map of cost-effective conservation across Europe. The red areas in Fig. \ref{fig:3a} are the most cost-effective while the blue areas are the least cost-effective. In this example, cost effective conservation policy for the EEA area as a whole would focus on the Iberian Peninsula and the Mediterranean, with Northern regions being a lesser priority. This spatial configuration reflects the fact that cost effectiveness stems from ratio of cost and the change in extinction risk (See p16 of SI Fig. S8 maps based solely on cost or marginal changes in persistence/extinction risk). Fig. \ref{fig:3b} shows the associated MBRC, colour coded in the same way as Fig. \ref{fig:3a}, with the most cost-effective (red) cells making up the initial segment of the curve transitioning to less cost-effective (blue) cells further along the curve. The shape of the curve illustrates a smoothly increasing marginal cost, reflecting both the heterogeneity in cost-effectiveness across the EEA, and the granular nature of the analysis. The SI (p11) explains the iterative approach taken to estimate the MBRC to account for the spatial dependence of extinction risk for species across the entire EEA as habitat is incrementally restored. Table \ref{tab:pricing-biodiversity} illustrates the shadow prices for different targets.

\subsection{Application of extinction risk based TCA to CBA}
Consider a project which converts 30km$^2$ of pristine forest within the EEA to a forest plantation enterprise (see Fig. S10 in the SI) with associated impacts on biodiversity. To evaluate the social cost of these biodiversity impacts the TCA approach has the following steps, the implications of which are shown in Table \ref{tab:pricing-biodiversity}. First, the target for biodiversity, here measured by persistence (1 - extinction risk), needs to be defined. Following \cite{Strassburg2020RestorationPriorities}, the target for persistence is measured in terms of a desired arithmetic average of the persistence index across the 1km grid cells in the EEA (Eq. 3 in Methods). At present average persistence for the EEA region is at 0.75 for the IUCN European Red List species considered in this example. Our central case is shown in the shaded upper panel of Table 1 in which we assume that the persistence function has z = 0.25. Column 1 shows four different and mutually exclusive possible targets for persistence that the administration might choose: the current level (no loss target) of 0.75, followed by three possible restoration targets that increase in their ambition for species persistence from 0.8 to 0.85 to 0.9. Each possible target for persistence has its own associated shadow price, measured for convenience in Table \ref{tab:pricing-biodiversity} as the cost of a 1 percentage point change in persistence (i.e. a 0.01 change, so Eq. 8 in Methods divided by 100). This cost increases with the ambition of the target as less and less cost effective land is required for restoration of suitable habitat, and we move up the MBRC curve in Fig. \ref{fig:3b} . Column 3 of Table 1 shows that the target of 0.75 (no loss) leads to an extremely small (nearly zero) shadow price of biodiversity. This reflects the extremely low cost of achieving marginal increases in persistence measured from the status quo of 0.75. The target of 0.9 has the highest shadow price of over \texteuro{}41bn.

\begin{table}[h]
\centering
\small

\newcolumntype{Y}{>{\centering\arraybackslash}X}

\begin{tabularx}{\linewidth}{YYYYY}
\toprule
\makecell{Target\\persistence} &
\makecell{$z$\\value} &
\makecell{MBRC:\\Shadow Price\\(€mn/percentage point)} &
\makecell{Plantation project\\impact on persistence\\(percentage points)} &
\makecell{Total cost of project\\impact on persistence\\(€mn)} \\
\midrule

\rowcolor{lightgreen}
0.75 & 0.25 & 0.07    & $-0.0017$ & $<0.01$ \\

\rowcolor{lightgreen}
0.80 & 0.25 & 1542.3  & $-0.0017$ & 2.6     \\

\rowcolor{lightgreen}
0.85 & 0.25 & 8839.8  & $-0.0017$ & 14.9    \\

\rowcolor{lightred}
0.90 & 0.25 & 41343.2 & $-0.0017$ & 69.5    \\
\hline

\midrule
\multicolumn{5}{c}{\textit{Sensitivity analysis (95\% confidence interval\cite{Strassburg2020RestorationPriorities})}} \\
\midrule

0.84 & 0.15 & 0.09 & $-0.0012$ & $<0.01$ \\ 

0.85 & 0.15 & 312.15 & $-0.0012$ & 0.37 \\ 
\rowcolor{lightred}
0.90 & 0.15 & 10661.38 & $-0.0012$ & 12.72 \\ 

\hline

0.68 & 0.35 & 0.07 & $-0.0020$ & $<0.01$ \\ 

0.80 & 0.35 & 6775.86 & $-0.0020$ & 13.55 \\ 

0.85 & 0.35 & 22111.65 & $-0.0020$ & 44.23 \\ 

\rowcolor{lightred}
0.90 & 0.35 & 307255.29 & $-0.0020$ & 614.60 \\ 

\hline

\bottomrule
\end{tabularx}

\caption{\textbf{Pricing biodiversity loss from habitat destruction}. CBA using target and cost-based prices for persistence of Red List Species in Europe (persistence $= 1 -$ extinction risk). Here, the impact on persistence is calculated at the target level of persistence (shown in column~1). Column~1 shows the different possible targets for persistence that an administration might choose, noting that the current average persistence for the EEA $= 0.75$ at the central $z$ value of $0.25$. Column~2 shows the different $z$ values for which this example was calculated. Column~3 shows the MBRC for each of the possible targets for persistence measured in €mn per percentage point: i.e., the marginal cost at different points (targets) on the MBRC curve in Figure \ref{fig:3b} in the main text. For example, the cost of increasing EEA average persistence from $0.89$ to $0.90$ (i.e., one percentage point) is €41{,}343\,mn, for $z = 0.25$. Column~4 shows the estimated impact of the plantation project on the metric of interest: average persistence across the EEA (in percentage points), with negative values showing reductions in persistence (increases in extinction risk) arising from the 900\,km$^{2}$ forest plantation project. The measured effect accounts for both positive and negative effects for the different species affected. Multiplying columns~3 and~4 together gives column~5, which shows the total shadow cost of the loss of persistence generated by the plantation project across different targets and hence differing MBRC for biodiversity. The sensitivity analysis shows the 95\% confidence intervals for the parameter $z$ used by \cite{Strassburg2020RestorationPriorities}}
\label{tab:pricing-biodiversity}

\end{table}

At the project level, column 4 of Table 1 measures the impact in percentage points on the EEA average persistence measures arising from the project intervention. The TCA approach multiplies this change (column 4) by the shadow price (column 3) to obtain the shadow cost of biodiversity (column 5). In this case the project induces a reduction in EEA persistence of 0.0017 percentage points and hence imposes a cost on meeting the EEA target. If the target is 0.9 then the cost is \texteuro{}69m, a figure that would appear as a cost in the social CBA of this project alongside other associated project costs and benefits. Fig. \ref{fig:3} and Table \ref{tab:pricing-biodiversity} illustrate the sensitivity of the shadow price to modelling assumptions. There is sensitivity to the stringency of the target and to the parameters of the persistence-habitat relationship. Fig. \ref{fig:3b} illustrates sensitivity using the 95\% confidence intervals for the elasticity $z$ in Eq. 1 in the Methods. This leads to values ranging from \texteuro{}13m to \texteuro{}615m around a central value of \texteuro{}70m for the forestry intervention when the target is to increase persistence (reduce extinction risk) to 0.9 (0.1). While the 95\% interval is a wide range for this parameter for aggregate analysis, the analysis makes quite clear that there is a trade-off between more stringent targets and the width of this interval, which would be reduced by more accurate and granular measures of the appropriate functional form for the persistence – habitat relationship.

\subsection{A species richness MBRC for the UK}

The EEA example illustrates how the TCA approach could be implemented at a cross-country scale using a widely accepted metric of biodiversity for which agreed targets exist. Here we present a national case study, using a high-resolution assessment of the MBRC in part of the UK, specifically England. The approach not only highlights the adaptability of the TCA approach across biodiversity metrics, but also shows how the MBRC can be adapted to incorporate different biodiversity-enhancing technologies as per Fig. \ref{fig:2}. In this illustration we change from the persistence - extinction risk metric of the previous example to the more widely applied average species richness measure, measured using the JNCC priority species data for the UK. The cost data reflects agricultural opportunity costs estimated using the NEV model developed by the LEEP Institution (see \cite{Day2024NaturalCapitalPolicies} and p20 of the SI for details).

\begin{figure}[h]
\centering

\begin{subfigure}{0.48\linewidth}
  \centering
  \includegraphics[width=\linewidth]{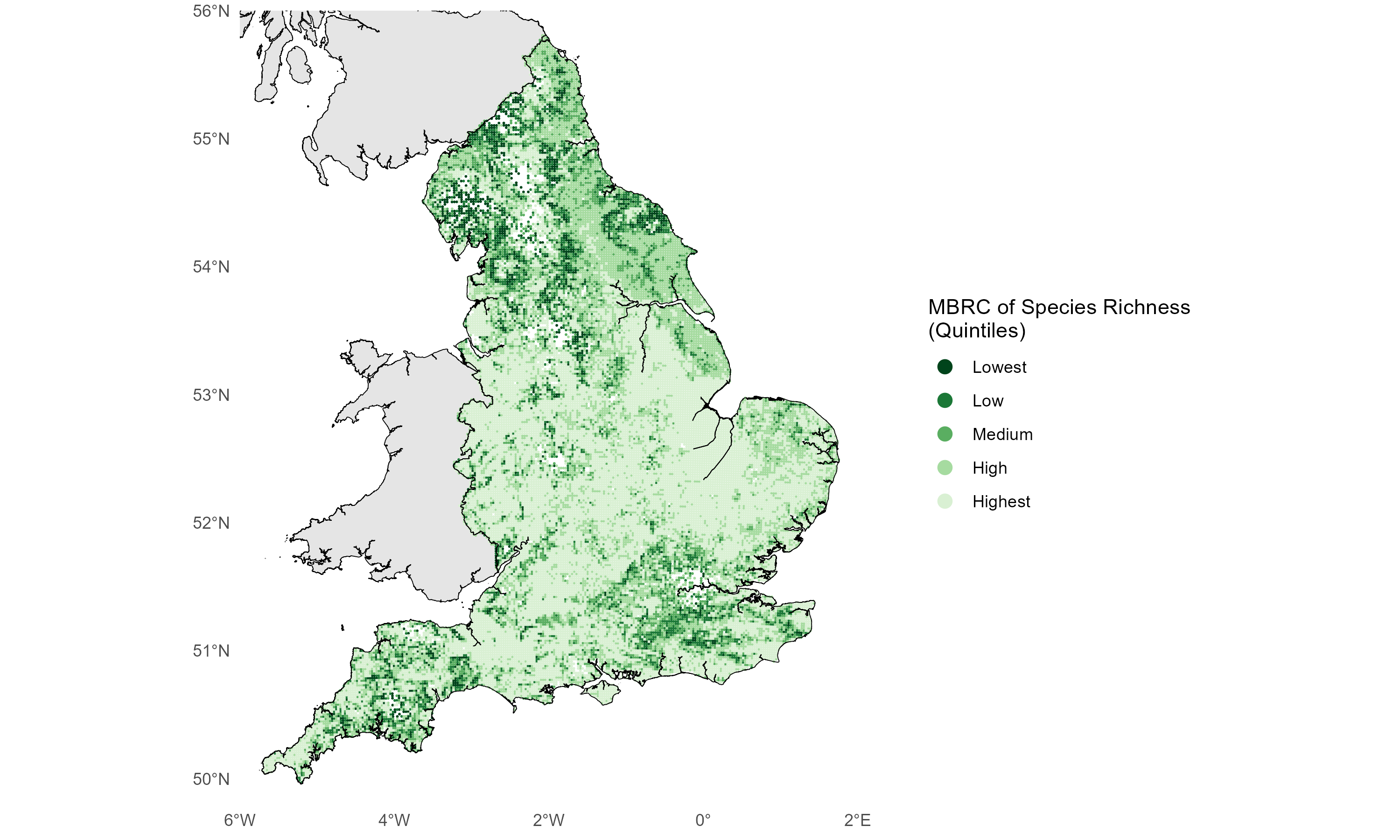}
  \caption{}
  \label{fig:4a}
\end{subfigure}\hfill%
\begin{subfigure}{0.48\linewidth}
  \centering
  \includegraphics[width=\linewidth]{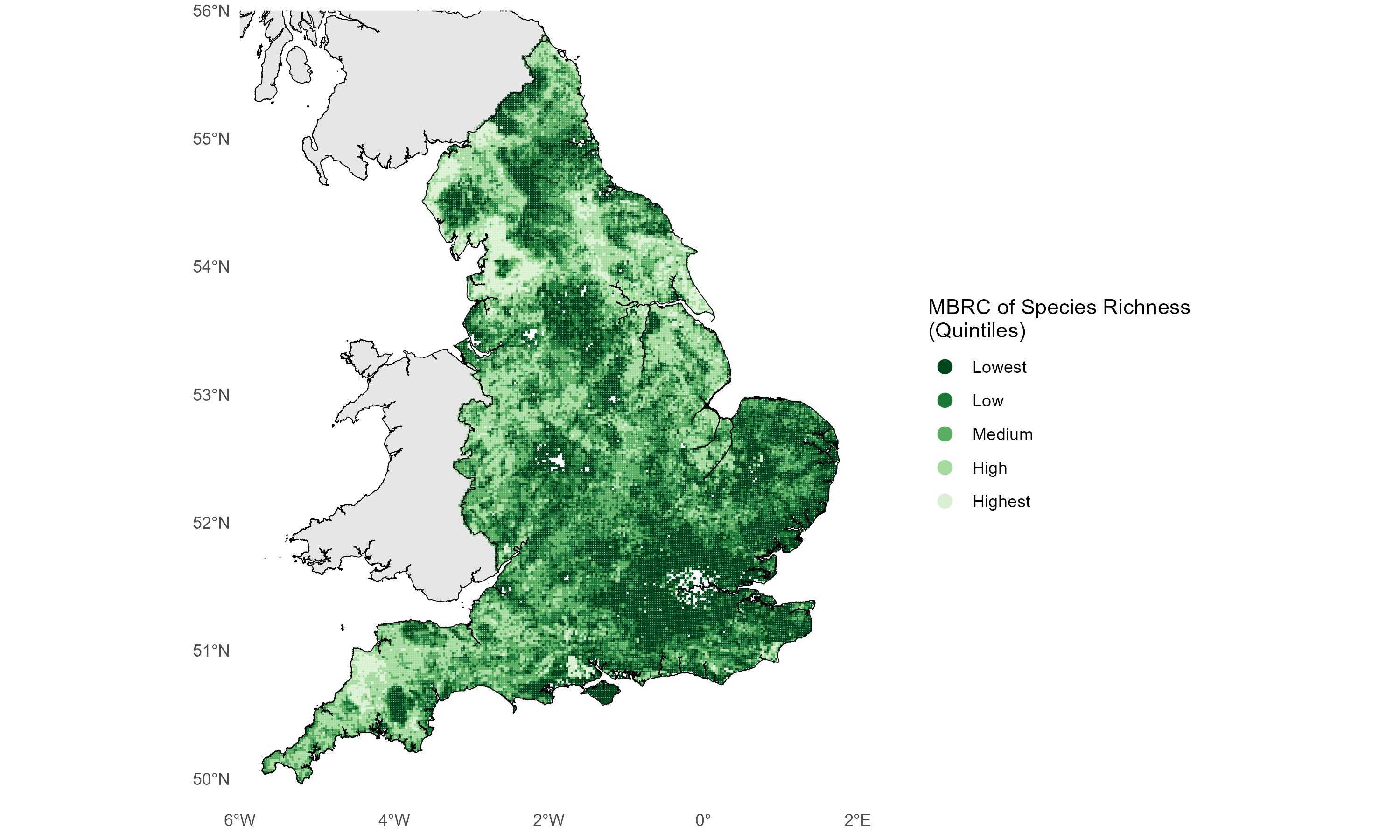}
  \caption{}
  \label{fig:4b}
\end{subfigure}
\caption{\textbf{Spatial and technological variation in the marginal biodiversity restoration cost (MBRC) across England}. Delivering the biodiversity associated with species-rich semi-natural grassland habitat by (a) conversion from arable farmland, or (b) destocking grassland. While the spatial pattern is relatively similar across technologies, restoration costs for technology (b) are generally (but not always) lower cost than for (a). Costs and biodiversity response were calculated using the land use models given by \cite{Day2024NaturalCapitalPolicies}.}
\label{fig:4}

\end{figure}

\begin{figure}[h]
\centering
\begin{subfigure}{0.48\linewidth}
  \centering
  \includegraphics[width=\linewidth]{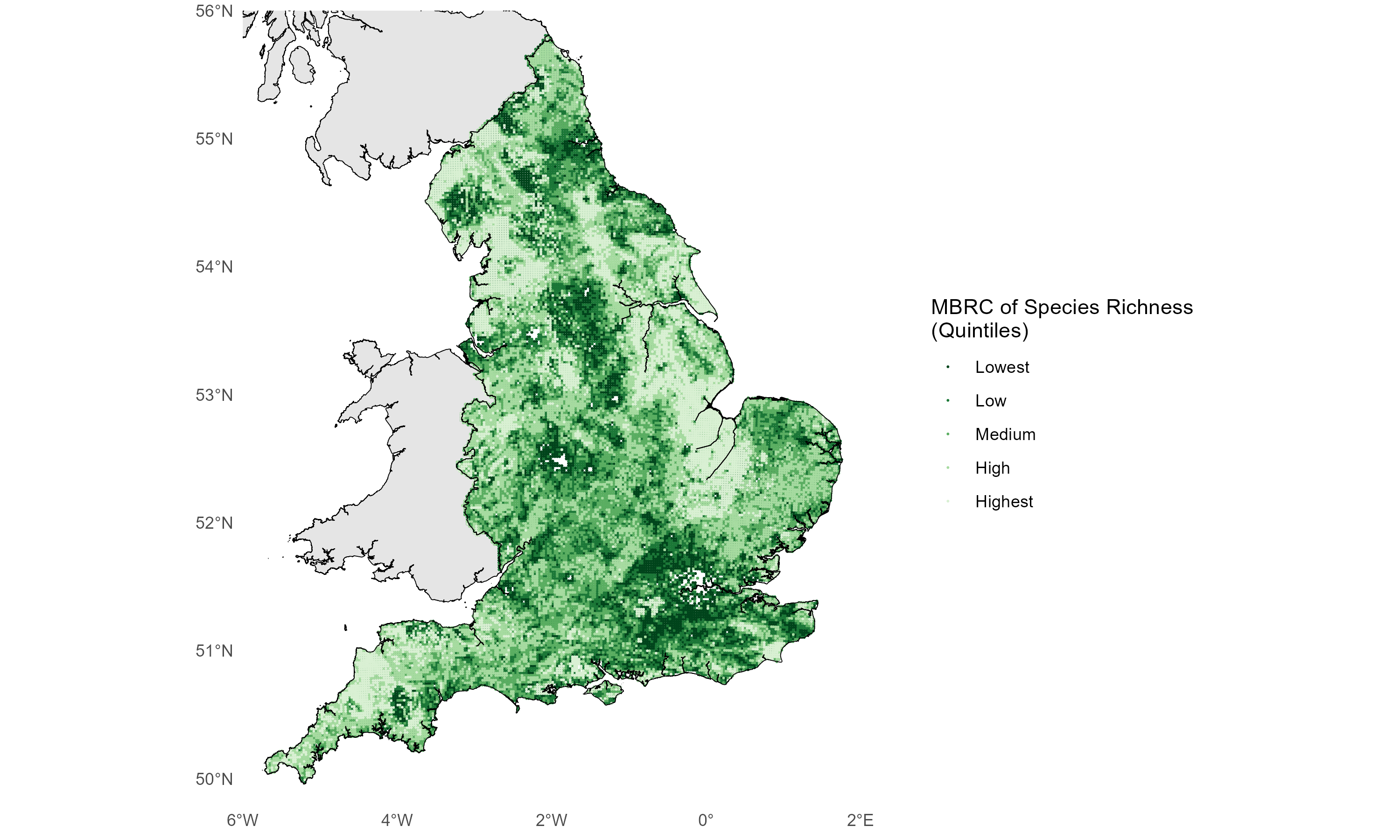}
  \caption{}
  \label{fig:5a}
\end{subfigure}\hfill%
\begin{subfigure}{0.48\linewidth}
  \centering
  \includegraphics[width=\linewidth]{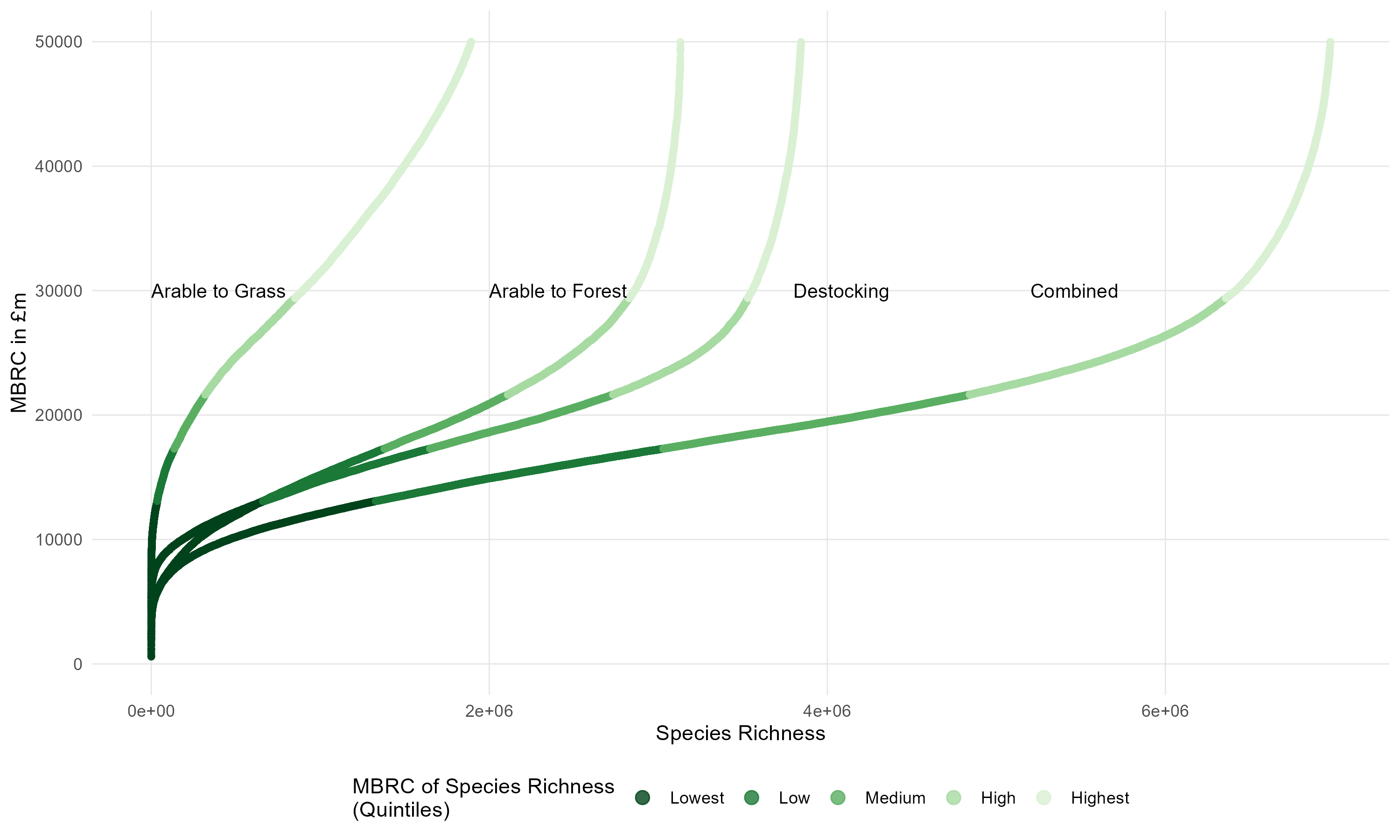}
    \caption{}
  \label{fig:5b}
\end{subfigure}
\caption{\textbf{An empirical Marginal Biodiversity Recovery Cost (MBRC) curve.} The cost curves are calculated across three `technologies': arable to species rich grassland; arable to forest/woodland; and livestock to species rich grassland (destocking). Separate MBRCs for each technology illustrate the heterogeneity of marginal costs (measured as opportunity costs of land). The `combined' MBRC curve reflects all three technologies ordered in terms of the cost of a percentage change in average species richness across the UK (see Methods). Costs and biodiversity response calculated using the land use models given by \cite{Day2024NaturalCapitalPolicies}.}
\label{fig:5}
\end{figure}

Fig. \ref{fig:4} illustrates the spatial variation in MBRC that arises from applying two alternative technologies to generate the same habitat outcome, semi-natural grassland, derived through either (a) converting arable farmland, or (b) removing livestock. Both approaches to biodiversity enhancement vary significantly across space, delivering both high and low cost-effectiveness. Yet in general, destocking is more cost effective than conversion of arable land. Fig. \ref{fig:5} extends this analysis by distinguishing between conversion of arable farmland to grassland and converting to forest/woodland. Fig. \ref{fig:5a} maps the resultant MBRC when the most cost-effective of these three technologies is applied (note the difference from the single technology maps of Fig. \ref{fig:4}). Fig. \ref{fig:5b} shows individual and combined MBRC curves. Initially conversion from arable to forest provides the most cost-effective approach, followed by converting arable to grassland. Destocking eventually becomes the cheapest technology as successive land use change takes place. The combined MBRC consistently takes the most cost-effective approach defining the optimal mix of technologies and locations for policy purposes. However, all technologies eventually reach areas of rapidly rising costs, reflecting both ecological constraints on improving species richness and the economic constraints of increasing opportunity cost of land. The shadow price can be obtained from selecting a specific target for species richness and reading across the associated value of the MBRC. This shadow price can be used CBA support for decision-making as described for the EEA study.

The TCA approach can be adapted to other biodiversity policy scenarios. Multiple metrics may be required to achieve qualitatively different targets for biodiversity (e.g. abundance, extinction risk, richness and habitat) and these can be incorporated into the estimation of metric-specific MBRCs. Issues such as habitat contiguity effects upon biodiversity can also be incorporated as can dynamic effects; the impact of changes in the timing of restoration upon MBRC curves. Furthermore, for dynamic cost efficiency, shadow prices can be expected to rise over time at the rate of discount (as in the carbon case) when considered ex ante and should be updated over time. Again, the price will reflect the stringency of the societal target, opportunity costs, and the biodiversity metric in which the policy targets are expressed.

The MBRC describes the relationship between changes in biodiversity and their corresponding costs and (as per current practice for the shadow price of carbon) does not include the value of any ecosystem services (or socio-political impacts) arising from restoration projects. Nevertheless, policies that affect outcomes for biodiversity typically also affect other ecosystem services such as carbon storage, recreation, water quality and flood risk. With the introduction of the TCA approach all of these can now be incorporated within CBA assessments, whether as shadow prices or via robust monetary valuation \cite{Dasgupta2021Biodiversity} allowing decision makers to discriminate between spending options without omitting either biodiversity or other benefits.

\section{Conclusion}

Reliance upon international treaties has failed to reverse global biodiversity loss. We argue that these global initiatives require support from the incorporation of biodiversity within routine national and regional decision-making. At present the absence of a shadow price leaves biodiversity invisible in conventional CBA and related economic decision-making. Building on current practice with respect to including greenhouse gas removal within economic appraisal, the Target and Cost Analysis (TCA) approach offers a tractable means of pricing biodiversity in the form of the Marginal Biodiversity Recovery Cost (MBRC) curve. This yields a target compatible shadow price, which brings biodiversity into government decision-making and could readily be applied within the private sector. Compared to other alternatives, application of the TCA approach should not introduce significant political economy or acceptability issues and indeed may reduce them (See p25 of SI). Furthermore, knowledge of the MBRC is informative for many other conservation purposes.

The TCA approach avoids the difficulties of applying conventional economic valuation methods to biodiversity. Instead, as in the case of carbon pricing, scientifically evidenced targets can be related to corresponding shadow prices using standard economic costing methods incorporating risks and uncertainties, for example those concerning the need to avoid the economic damage arising from ecosystem service collapse (e.g. \cite{FrankSudarshan2024Vultures,Frank2024PestControl,ManningAndo2022BatServices}).

While the TCA parallels the MAC approach to shadow pricing carbon, there are important differences in the application to biodiversity. For instance, while greenhouse gases can be readily translated into a single fungible metric, this is not the case for biodiversity where various measures reflect different aspects of biodiversity and nature. However, many of these challenges are common to any biodiversity assessment, while the development of officially recognised metrics and targets provides a way forward.

A further difference between carbon and biodiversity shadow pricing concerns the responsibility for ensuring wildlife is enhanced. Within the UK, the 2019 amendment to the Climate Change Act 2008 \cite{UKGov2019ClimateActAmendment} makes it the Government's responsibility to attain the target of net zero emissions of greenhouse gases by 2050. This provides an efficient approach to tackling climate change with CBA assessments of public spending projects required to include the MAC-derived shadow price for carbon across all investments. However, in the carbon case those investments do not have to undertake individual offset projects to physically offset for any emissions they cause; that responsibility falls upon the Government as part of its net zero commitment. A comparable commitment to delivering biodiversity targets is necessary to facilitate general application of the TCA approach. While the 2021 Environment Act \cite{UKGov2021EnvironmentAct} provides such a general commitment, to date its BNG rules limit implementation to development and infrastructure projects, requiring developers to fund offsets rather than making meeting biodiversity commitments a government responsibility. Comparing the success of the carbon pricing rules with the international track record of failure to meet biodiversity targets (e.g. many of the Aichi targets of the CBD were not met), the introduction of shadow prices for biodiversity, backed by a government commitment to deliver corresponding targets, would offer much needed new potential for change and address the mainstreaming requirements of the Kunming-Montreal Agreement.

A final difference between the carbon and biodiversity TCA approaches concerns spatial variation. While individuals may care about the location of the physical infrastructure associated with carbon emissions or removal, because greenhouse gases are generally invisible and mix perfectly in the atmosphere, their contribution to climate change does not vary according to the location of emissions or removals; only their warming potential matters. This is not the case for biodiversity where a change in the location of losses or gains typically alters the species affected and impact generated. This is a particular problem where the substitutability of one biodiversity component for another is imperfect. Where a component is entirely spatially non-fungible then, where such replacements are not non-substitutable and regulated in other ways (e.g. for ancient woodland), a like-for-like replacement cost version of the biodiversity TCA approach is required. Indeed, the UK Government considers this as a first step for including biodiversity shadow pricing in the HM Treasury Green Book rules for the appraisal of public policies, programmes and projects. In the absence of like-for-like, the mitigation hierarchy would bind, or else replacement costs would be come prohibitively expensive to the same effect: avoidance of biodiversity loss.

\section{Materials and Methods}

\subsection{MBRC curve for a biodiversity metric: a general approach}

Cost-effective prioritisation of habitat restoration to increase biodiversity in Europe has several requirements. First, measurement of persistence across the EEA and marginal costs of land use change need to be calculated. In our illustrative application biodiversity is measured as species' regional likelihood of persistence (which is $1$ minus extinction risk). Europe is chosen as an illustrative study area because a consistent opportunity cost layer of restoration is available for $37$ contiguous European countries (e.g. \cite{Spencer2024GreenInfrastructure}) and because the policies and regional cooperation of the European Union and European Environmental Agency render the research particularly relevant in this region.

The theory connecting habitat and extinction risk stems from species--area relationships. With $S(H)$ representing the number of species expected to persist in the long run given an area of habitat $H$, $z$ a parameter between $0$ and $1$ specifying sensitivity of $S$ to the proportion of remaining habitat $H$, and $\gamma$ a constant reflecting heterogeneity of the ecosystem, the relationship is typically written as $S(H)=\gamma H^{z}$. With original (pre-human disturbance) habitat given by $OH$ the original number of species would be $S(OH)=\gamma (OH)^{z}$, and the proportion of species remaining is then given by $S(H)/S(OH)=(H/OH)^{z}$. This relationship is often applied at different spatial scales: regional, global etc., but also can be applied at the level of a species $i$. Under the assumption of independence of species survival, $(H/OH)^{z}$ can be interpreted as the probability that a species still persists, $p_i$, given the proportion of remaining habitat $H_i/OH_i$. The formula for persistence for a given species $i$ is given by:

\begin{equation}
p_i=\left(\frac{H_i}{OH_i}\right)^z
\end{equation}

Extinction risk is then given by $1-p_i$. While this measure is not necessarily perfect, it is transparent, frequently used, and captures the habitat related determinants of extinction risk \cite{Strassburg2020RestorationPriorities}. It also makes clear the relevance of existing habitat, $H_i$, as a policy variable for reducing extinction risk.

The power $z$ governs the sensitivity of the probability of persistence to the proportion of habitat remaining. A higher value of $z$ implies greater sensitivity to habitat loss. Following prior studies, $z = 0.25$ is used as the central estimate \cite{duran_practical_2020,Strassburg2020RestorationPriorities,arrhenius_species_1921}, with sensitivity in the range z = 0.15 to 0.35, the 95\% confidence interval used in \cite{Strassburg2020RestorationPriorities}. Overall, this theory reflects the idea that remaining habitat is akin to an isolated island within a larger area of habitat destroyed through human intervention. This provides a snapshot in time of the probability of long-run species persistence. For ease of exposition, we ignore contiguity through proximity, although this could be incorporated in more detailed specifications.

The arithmetic mean of $p_i$ across species $i = 1, ., N.$ provides a biodiversity index, $I$, that reflects the proportion of species expected to persist in the long run, given current habitat conditions:

\begin{equation}
I=\frac{1}{N}\sum_{i=1}^{N}p_i=\frac{1}{N}\sum_{i=1}^{N}\left(\frac{H_i}{OH_i}\right)^z
\end{equation}

We calculate the index $I$ for the EEA and present the biodiversity target in terms of a desired future value of $I$. For a single species i the marginal benefit ($MB_i$) of an additional unit of habitat is given by:

\begin{equation}
MB_i=\frac{\partial p_i}{\partial H_i}=z\,(H_i)^{z-1}\left(\frac{H_i}{OH_i}\right)^z
\end{equation}

The marginal impact of restoring a unit of land (cell $j$) across species is expressed as:

\begin{equation}
MB_j=\frac{1}{N}\sum_{i=1}^{N}\left(\frac{\partial p_i}{\partial H_i}\right)\left(\frac{\partial H_i}{\partial R_j}\right)
\end{equation}

\noindent which can be understood as the mean of species $i$ - specific marginal benefits. Here,  $\partial H_i/\partial R_j$ reflects the effect of restoration area in cell $j$ ($R_j$) on the habitat for species $i$ ($H_i$) which equals 1 if species $i$ gains a unit of suitable habitat, minus 1 if it loses a unit of suitable habitat, and equals 0 for species that are not in range in cell $j$ or for which neither the previous nor the restored habitat is suitable. Note that if $\partial H_i/\partial R_j=-1$ this indicates that the previous habitat was suitable for species $i$ but the restored habitat is not, e.g. for species that prefer arable land but not forest, where forest would emerge if arable land were restored.

Given the cost of intervention $c_j$, in principle reflecting opportunity costs, restoration costs and other transactions costs, the cost effectiveness (CE) of biodiversity gains from restoring cell $j$ is:

\begin{equation}
CE_j=\frac{MB_j}{c_j}
\end{equation}

\noindent Grid squaresare then ordered in terms of their cost effectiveness as in Fig \ref{fig:3a}. The inverse of this relationship yields the MBRC curve in Fig \ref{fig:3b}:

\begin{equation}
MBRC=\frac{c_j}{MB_j}
\end{equation}

\noindent which is the marginal cost of changing the biodiversity index, here persistence, by 1 unit. A change from 0 to 1 in persistence means removing extinction risk entirely. In Table \ref{tab:pricing-biodiversity} we present the MBRC values more intuitively in terms of percentage points.

\subsection{Data}

Data on the opportunity cost of land were sourced from \cite{Spencer2024GreenInfrastructure} for 37 member countries of the EIONET collaborative network of the European Environment Agency (as of 2019 and including the UK)\cite{eea_copernicus_2019}. A 1km x 1km resolution grid was used to prepare the data, with restoration decisions made at a 5km x 5km resolution. Land prices, which reflect the present value of the stream of profits expected from the use of land, are not available at the granularity required for the analysis. Instead, we use data on observed annual land rents from \cite{spencer_spatial_2024} and transform it to an asset value (the present value of the stream of rents over time discounted at a rate of 5\%) to proxy land purchase prices. Active restoration costs and co-benefits are not included but would be valuable in future assessments, although land rents reflect the major proportion of costs in most cases\cite{Bateman2015MarketForcesSpatialTargeting}.

Current habitat and restored habitat maps are provided by \cite{jung_layer_2020}, where restored habitat is assumed to take its original habitat, prior to human intervention. These maps use land cover, climate, and topography data. Habitat is classified using Level 1 IUCN habitat types, with 'artificial' habitat classified using Level 2. Restoration assumes changing from current use to original land use cover, which often means pristine ecosystems or as defined by\cite{Hengl2018PotentialNaturalVegetation} on p3: ``the hypothetical vegetation cover that would be present if the vegetation were in equilibrium with environmental controls, including climatic factors and disturbance, and not subject to human management''. Species range data were taken from the IUCN Red List for Europe \cite{IUCN2024EuropeanRedList, iucn_red_list_2024 }.

\subsection{Empirical Approach}

Species range data, habitat and elevation preferences, and maps of current and potential habitat (the latter of which represents a hypothetical scenario of full natural habitat cover) were used to create area of habitat maps for terrestrial mammals, amphibians, and reptiles on the European Red List\cite{iucn_habitats_2024}. Habitat suitability and elevation preference data were obtained from the IUCN\cite{iucn_habitats_2024,hanson_aoh_2024}. For species with complete preference information, areas marked as ``present'' and located in Europe were selected from the range maps, which were also obtained from\cite{iucn_spatial_2024}. These areas were aggregated to produce one regional range map per species.

Based on these processed range maps, two maps were generated for each species: one based on the current habitat map ($H$), and one based on the potential habitat map ($OH$). For each species and habitat map, areas of the processed range map that (i) fall inside the study area of 37 EIONET countries, (ii) have a suitable habitat type, and (iii) lie within the species' preferred elevation range were identified. When evaluated against the current habitat map, these areas represent the species' present extent of suitable habitat; when evaluated against the potential habitat map, these areas represent the extent under fully natural habitat, which serves as the baseline and informs how a species' area of habitat changes ($H$ changes) as parts of the study area are restored.

Of the 526 species of terrestrial mammals, amphibians, and reptiles extracted from the European Red List, not all are included in the analysis. Species lacking complete habitat and elevation preference data, or those outside the study area, were excluded. The final sample includes 218 species.

The UK example follows a similar process to account for average species richness. See p 21 of the SI for more details of the specific approach taken there.

\bibliography{NatSustBib.bib}

@book{Dasgupta2021Biodiversity,
  author  = {Dasgupta, Partha},
  year    = {2021},
  title   = {The economics of biodiversity: The Dasgupta review},
  publisher = {HM Treasury},
  address = {London}
}

@article{Steffen2015PlanetaryBoundaries,
  author  = {Steffen, Will and Richardson, Katherine and Rockstr{\"o}m, Johan and Cornell, Sarah E. and others},
  year    = {2015},
  title   = {Planetary boundaries: Guiding human development on a changing planet},
  journal = {Science},
  volume  = {347},
  number  = {6223},
  doi     = {10.1126/science.1259855}
}

@techreport{ipbes_global_2019,
  address     = {Bonn, Germany},
  title       = {Global assessment report on biodiversity and ecosystem services of the {Intergovernmental} {Science}-{Policy} {Platform} on {Biodiversity} and {Ecosystem} {Services}},
  institution = {IPBES secretariat},
  author      = {{IPBES}},
  editor      = {Brondízio, Eduardo Sonnewend and Settele, Josef and Diaz, Sandra and Ngo, Hien Thu},
  month       = {may},
  year        = {2019},
  doi         = {10.5281/zenodo.6417333}
}

@incollection{IPCC2023AR6SynthesisSections,
  author    = {{IPCC}},
  year      = {2023},
  title     = {Sections},
  booktitle = {Climate change 2023: Synthesis report. Contribution of Working Groups I, II and III to the Sixth Assessment Report of the Intergovernmental Panel on Climate Change},
  editor    = {Lee, Hoesung and Romero, Jose},
  publisher = {IPCC},
  address   = {Geneva, Switzerland},
  pages     = {35--115},
  doi       = {10.59327/IPCC/AR6-9789291691647}
}

@misc{WorldBank2024CarbonShadowPrice,
  author      = {{World Bank}},
  year        = {2024},
  title       = {Guidance note on shadow price of carbon in economic analysis},
  institution = {World Bank Group},
  address     = {Washington, DC},
  url         = {http://documents.worldbank.org/curated/en/099553203142424068}
}

@article{Withey2012ConservationROI,
  author  = {Withey, Jeffrey C. and Lawler, Joshua J. and Polasky, Stephen and Plantinga, Andrew J. and Nelson, Erik J. and Kareiva, Peter and Wilsey, Christopher B. and Schloss, Cynthia A. and Nogeire, Tina M. and Ruesch, Aaron and Ramos, Jose Jr. and Reid, Walter},
  year    = {2012},
  title   = {Maximising return on conservation investment in the conterminous {USA}},
  journal = {Ecology Letters},
  volume  = {15},
  number  = {9},
  pages   = {1040--1048},
  doi     = {10.1111/j.1461-0248.2012.01847.x}
}

@article{Ando1998SpeciesDistributions,
  author  = {Ando, Amy and Camm, Jeffrey and Polasky, Stephen and Solow, Andrew},
  year    = {1998},
  title   = {Species distributions, land values, and efficient conservation},
  journal = {Science},
  volume  = {279},
  pages   = {2126--2128}
}

@misc{Natura2000Overview,
  author = {{European Commission}},
  title  = {Natura 2000 overview},
  url    = {https://environment.ec.europa.eu/topics/nature-and-biodiversity/natura-2000_en}
}

@misc{UKGov2023EnvironmentalImprovementPlan,
  author      = {{HM Government}},
  year        = {2023},
  title       = {Environmental improvement plan 2023},
  institution = {UK Government},
  url         = {https://www.gov.uk/government/publications/environmental-improvement-plan-2023}
}

@article{Strassburg2020RestorationPriorities,
  author  = {Strassburg, Bernardo B. N. and others},
  year    = {2020},
  title   = {Global priority areas for ecosystem restoration},
  journal = {Nature},
  volume  = {586},
  number  = {7831},
  pages   = {724--729},
  doi     = {10.1038/s41586-020-2784-9}
}

@misc{Spencer2024GreenInfrastructure,
  author      = {Spencer, David and Marques, Ana and Veerkamp, Casper and van der Marel, Maarten and Verburg, Peter and Namasivayam, Anusha S. and Di Marco, Moreno and Jung, Martin and Kujala, Heini and O'Connor, Lauren and Visconti, Piero and Schipper, Aafke},
  year        = {2024},
  title       = {Spatial opportunities and constraints for green infrastructure network design},
  institution = {Project report},
  doi         = {10.3897/arphapreprints.e123365}
}

@misc{jung_layer_2020,
	title = {A layer of global potential habitats},
	url = {https://zenodo.org/records/4038749},
	doi = {10.5281/zenodo.4038749},
	language = {eng},
	urldate = {2024-07-30},
	publisher = {Zenodo},
	author = {Jung, Martin},
	month = sep,
	year = {2020},
}

@misc{hanson_aoh_2024,
  author       = {Hanson, Jeffrey O.},
  title        = {{aoh}: Create Area of Habitat Data},
  year         = {2024},
  version      = {1.0.0},
  publisher    = {GitHub},
  url          = {https://github.com/prioritizr/aoh},
  note         = {R package. Accessed 2024}
}

@misc{spencer_spatial_2024,
	type = {Project {Report}},
	series = {{NaturaConnect} - {Designing} a resilient and coherent {TransEuropean} {Network} for {Nature} and {People}},
	title = {Spatial opportunities and constraints for green infrastructure network design},
	copyright = {http://creativecommons.org/licenses/by/4.0/},
	doi = {10.3897/arphapreprints.e123365},
	language = {en},
	urldate = {2024-07-30},
	author = {Spencer, Douglas and Marques, Alexandra and Veerkamp, Clara and Van Der Marel, Martijn and Verburg, Peter and Namasivayam, Anandi Sarita and Di Marco, Moreno and Jung, Martin and Kujala, Heini and O'Connor, Louise and Visconti, Piero and Schipper, Aafke},
	month = mar,
	year = {2024},
}

@misc{eea_copernicus_2019,
	title = {Copernicus {EEA39} {Boundary} {Layer} with 250 m buffer (raster 100m)},
	url = {https://sdi.eea.europa.eu/catalogue/srv/api/records/8a1b3765-a277-4505-a316-28bbe53a00b5},
	author = {{EEA}},
	month = jan,
	year = {2019},
}

@article{arrhenius_species_1921,
	title = {Species and {Area}},
	volume = {9},
	issn = {0022-0477},
	url = {https://www.jstor.org/stable/2255763},
	doi = {10.2307/2255763},
	number = {1},
	urldate = {2025-02-28},
	journal = {Journal of Ecology},
	author = {Arrhenius, Olof},
	year = {1921},
	note = {Publisher: [Wiley, British Ecological Society]},
	pages = {95--99},
}

@article{Hengl2018PotentialNaturalVegetation,
  author  = {Hengl, Tomislav and Walsh, Markus G. and Sanderman, Jonathan and Wheeler, Ian and Harrison, Sandy P. and Prentice, I. Colin},
  year    = {2018},
  title   = {Global mapping of potential natural vegetation: An assessment of machine learning algorithms for estimating land potential},
  journal = {PeerJ},
  volume  = {6},
  pages   = {e5457},
  doi     = {10.7717/peerj.5457},
  url     = {https://doi.org/10.7717/peerj.5457}
}

@misc{iucn_spatial_2024,
	title = {Spatial {Data} {Download}},
	url = {https://www.iucnredlist.org/resources/spatial-data-download},
	urldate = {2024-06-15},
	journal = {IUCN Red List of Threatened Species},
	author = {{IUCN}},
	year = {2024},
    note = {https://www.iucnredlist.org/resources/spatial-data-download, Accessed 2024}
}

@misc{jung_global_2020,
	title = {A global map of terrestrial habitat types},
	doi = {10.5281/zenodo.4058819},
	language = {eng},
	urldate = {2024-07-30},
	publisher = {Zenodo},
	author = {Jung, Martin and Dahal, Prabhat Raj and Butchart, Stuart H. M. and Donald, Paul F. and De Lamo, Xavier and Lesiv, Myroslava and Kapos, Valerie and Rondinini, Carlo and Visconti, Piero},
	month = sep,
	year = {2020},
}

@article{jung_assessment_2024,
	title = {An assessment of the state of conservation planning in {Europe}},
	volume = {379},
	url = {https://doi.org/10.1098/rstb.2023.0015},
	language = {en},
	journal = {Philosophical Transactions B},
	author = {Jung, Martin and Alagador, Diogo and Chapman, Melissa and Hermoso, Virgilio and Kujala, Heini and Visconti, Piero},
	year = {2024},
	pages = {20230015},
}

@misc{iucn_habitats_2024,
	title = {Habitats {Classification} {Scheme} ({Version} 3.1)},
	url = {https://www.iucnredlist.org/resources/habitat-classification-scheme},
	journal = {IUCN Red List of Threatened Species},
	author = {{IUCN}},
	year = {2024},
    note = {https://www.iucnredlist.org/resources/habitat-classification-scheme. Accessed 2024} 
}

@misc{IUCN2024EuropeanRedList,
  author       = {{International Union for Conservation of Nature}},
  title        = {European Red List},
  year         = {2024},
  publisher    = {IUCN},
  url          = {https://www.iucnredlist.org/regions/europe},
  note         = {https://www.iucnredlist.org/regions/europe, Accessed 2024}
}

@misc{iucn_red_list_2024,
  author       = {{International Union for Conservation of Nature}},
  title        = {The IUCN Red List of Threatened Species},
  year         = {2024},
  version      = {2024-1},
  publisher    = {IUCN},
  url          = {https://www.iucnredlist.org},
  note         = {https://www.iucnredlist.org. Accessed 2024}
}

@article{Day2024NaturalCapitalPolicies,
  author  = {Day, Brett and Mancini, Matteo and Bateman, Ian J. and Binner, Amy and Cho, Francesca and de Gol, Alexander and Smith, Graham},
  year    = {2024},
  title   = {Natural capital approaches for the optimal design of policies for nature recovery},
  journal = {Philosophical Transactions of the Royal Society B},
  volume  = {379},
  number  = {1903},
  pages   = {20220327},
  doi     = {10.1098/rstb.2022.0327}
}

@article{FrankSudarshan2024Vultures,
  author  = {Frank, Eyal and Sudarshan, Anant},
  year    = {2024},
  title   = {The social costs of keystone species collapse: Evidence from the decline of vultures in {India}},
  journal = {American Economic Review},
  volume  = {114},
  number  = {10},
  pages   = {3007--3040},
  doi     = {10.1257/aer.20230016}
}

@article{kesicki_marginal_2012,
  title      = {Marginal abatement cost curves: a call for caution},
  volume     = {12},
  issn       = {1469-3062, 1752-7457},
  shorttitle = {Marginal abatement cost curves},
  doi        = {10.1080/14693062.2011.582347},
  language   = {en},
  number     = {2},
  urldate    = {2024-07-30},
  journal    = {Climate Policy},
  author     = {Kesicki, Fabian and Ekins, Paul},
  year       = {2012},
  pages      = {219--236}
}

@misc{UKGov2019ClimateActAmendment,
  author       = {{HM Government}},
  year         = {2019},
  title        = {The Climate Change Act 2008 (2050 target amendment) order 2019},
  howpublished = {Legislation.gov.uk},
  url          = {https://www.legislation.gov.uk/ukdsi/2019/9780111187654},
  note         = {Accessed 2025-12-17}
}

@misc{DESNZ2024EnergyGHGValuation,
  author      = {{Department for Energy Security and Net Zero}},
  year        = {2024},
  title       = {Valuation of energy use and greenhouse gas (GHG) emissions: Supplementary guidance to the Green Book on appraisal and evaluation in central government},
  institution = {UK Government},
  url         = {https://assets.publishing.service.gov.uk/media/65aadd020ff90c000f955f17/valuation-of-energy-use-and-greenhouse-gas-emissions-for-appraisal.pdf}
}

@misc{HMTreasury2021DasguptaResponse,
  author      = {{HM Treasury}},
  year        = {2021},
  title       = {The economics of biodiversity: The Dasgupta Review -- government response},
  institution = {UK Government},
  url         = {https://www.gov.uk/government/publications/the-economics-of-biodiversity-the-dasgupta-review-government-response}
}

@misc{Robinson2019BCAGuidelines,
  author      = {Robinson, Lisa A. and Hammitt, James K. and Cecchini, Marco and Chalkidou, Kalipso and Claxton, Karl and Cropper, Maureen and Eozenou, Patrick H.-V. and de Ferranti, David and Deolalikar, Anil B. and Guanais, Frederico and Jamison, Dean T. and Kwon, Soonman and Lauer, Jeremy A. and O'Keeffe, Linda and Walker, Damian and Whittington, Dale and Wilkinson, Tom and Wilson, David and Wong, Bonnie},
  year        = {2019},
  title       = {Reference case guidelines for benefit-cost analysis in global health and development},
  institution = {Harvard T.H. Chan School of Public Health},
  address     = {Boston},
  url         = {https://content.sph.harvard.edu/wwwhsph/sites/2447/2019/05/BCA-Guidelines-May-2019.pdf}
}

@misc{EC2014CBAGuide,
  author      = {{European Commission}},
  year        = {2014},
  title       = {Guide to cost-benefit analysis of investment projects},
  institution = {European Commission, Directorate-General for Regional and Urban Policy},
  address     = {Brussels},
  url         = {https://ec.europa.eu/regional_policy/sources/studies/cba_guide.pdf}
}

@article{jetz_include_2022,
	title = {Include biodiversity representation indicators in area-based conservation targets},
	volume = {6},
	issn = {2397-334X},
	doi = {10.1038/s41559-021-01620-y},
	language = {en},
	urldate = {2024-07-30},
	journal = {Nature Ecology \& Evolution},
	author = {Jetz, Walter and McGowan, Jennifer and Rinnan, D. Scott and Possingham, Hugh P. and Visconti, Piero and O’Donnell, Brian and Londoño-Murcia, Maria Cecilia},
	month = feb,
	year = {2022},
	pages = {123--126},
}

@article{Keck2025HumanImpact,
  author  = {Keck, Fr{\'e}d{\'e}ric and others},
  year    = {2025},
  title   = {The global human impact on biodiversity},
  journal = {Nature},
  volume  = {641},
  number  = {8062},
  pages   = {395--400}
}

@article{Mancini2024Offsets,
  author  = {Mancini, Matteo C. and Collins, Rachel M. and Addicott, Emily T. and Balmford, Ben J. and Binner, Amy and Bull, Joseph W. and Day, Brett H. and Eigenbrod, Felix and zu Ermgassen, Sophie O. S. E. and Faccioli, Martina and Fezzi, Carlo and Groom, Ben and Milner-Gulland, E. J. and Owen, Neil and Tingley, Daniel and Wright, Emma and Bateman, Ian J.},
  year    = {2024},
  title   = {Biodiversity offsets perform poorly for both people and nature, but better approaches are available},
  journal = {One Earth},
  volume  = {7},
  pages   = {2165--2174}
}

@article{Guilhaumon2008SARUncertainty,
  author  = {Guilhaumon, Fran{\c{c}}ois and Gimenez, Olivier and Gaston, Kevin J. and Mouillot, David},
  year    = {2008},
  title   = {Taxonomic and regional uncertainty in species-area relationships and the identification of richness hotspots},
  journal = {Proceedings of the National Academy of Sciences of the United States of America},
  volume  = {105},
  number  = {40},
  pages   = {15458--15463},
  doi     = {10.1073/pnas.0803610105}
}

@article{Bateman2015MarketForcesSpatialTargeting,
  author  = {Bateman, I. J. and Coombes, E. and Fitzherbert, E. and Badura, T. and Binner, A. and Carbone, C. and Fisher, B. and Naidoo, R. and Watkinson, A. R.},
  year    = {2015},
  title   = {Conserving tropical biodiversity via market forces and spatial targeting},
  journal = {Proceedings of the National Academy of Sciences of the United States of America},
  volume  = {112},
  number  = {24},
  pages   = {7408--7413},
  doi     = {10.1073/pnas.1406484112}
}

@misc{MingWilliamson2023LinkedDeepGP,
  author       = {Ming, D. and Williamson, D.},
  year         = {2023},
  title        = {Linked Deep Gaussian Process Emulation for Model Networks},
  howpublished = {arXiv preprint},
  eprint       = {2306.01212},
  archivePrefix = {arXiv}
}

@article{MingWilliamsonGuillas2023DeepGPImputation,
  author  = {Ming, D. and Williamson, D. and Guillas, S.},
  year    = {2023},
  title   = {Deep Gaussian process emulation using stochastic imputation},
  journal = {Technometrics},
  volume  = {65},
  number  = {2},
  pages   = {150--161}
}

@article{BodeMurdoch2009SARRobustness,
  author  = {Bode, Michael and Murdoch, William},
  year    = {2009},
  title   = {Cost-effective conservation decisions are robust to uncertainty in the species-area relationship},
  journal = {Proceedings of the National Academy of Sciences of the United States of America},
  volume  = {106},
  number  = {7},
  pages   = {E12--E13},
  doi     = {10.1073/pnas.0811568106}
}

@article{Arlidge2018MitigationHierarchy,
  author  = {Arlidge, William N. S. and Bull, Joseph W. and Addison, Prue F. E. and Burgass, Michael J. and Gianuca, Dimas and Gorham, Taylor M. and Jacob, C{\'e}line and Shumway, Nicole and Sinclair, Samuel P. and Watson, James E. M. and Wilcox, Chris and Milner-Gulland, E. J.},
  year    = {2018},
  title   = {A global mitigation hierarchy for nature conservation},
  journal = {BioScience},
  volume  = {68},
  number  = {5},
  pages   = {336--347},
  doi     = {10.1093/biosci/biy029}
}

@article{Weitzman1992OnDiversity,
  author  = {Weitzman, Martin L.},
  year    = {1992},
  title   = {On diversity},
  journal = {The Quarterly Journal of Economics},
  volume  = {107},
  number  = {2},
  pages   = {363--405}
}

@article{Akcakaya2018IUCNGreenList,
  author  = {Ak{\c{c}}akaya, H. Resit and Bennett, Elizabeth L. and Brooks, Thomas M. and Grace, Michael K. and others},
  year    = {2018},
  title   = {Quantifying species recovery and conservation success to develop an {IUCN} Green List of species},
  journal = {Conservation Biology},
  volume  = {32},
  number  = {5},
  pages   = {1128--1138},
  doi     = {10.1111/cobi.13112}
}

@article{Mace2008IUCNExtinctionRisk,
  author  = {Mace, Georgina M. and Collar, Nigel J. and Gaston, Kevin J. and Hilton-Taylor, Craig and others},
  year    = {2008},
  title   = {Quantification of extinction risk: {IUCN}'s system for classifying threatened species},
  journal = {Conservation Biology},
  volume  = {22},
  number  = {6},
  pages   = {1424--1442},
  doi     = {10.1111/j.1523-1739.2008.01044.x}
}

@misc{UKGovStatutorySpeciesTargets,
  author      = {{UK Government}},
  year        = {2024},
  title       = {Statutory species targets: Environment Act target delivery plan},
  institution = {UK Government},
  url         = {https://www.gov.uk/government/publications/statutory-species-targets-environment-act-target-delivery-plan/statutory-species-targets-environment-act-target-delivery-plan}
}

@misc{DEFRA2018EnvironmentPlan,
  author      = {{Department for Environment, Food and Rural Affairs}},
  year        = {2018},
  title       = {A green future: Our 25 year plan to improve the environment},
  institution = {UK Government},
  url         = {https://www.gov.uk/government/publications/25-year-environment-plan}
}

@article{Pindyck2013JEL,
  author  = {Pindyck, Robert S.},
  year    = {2013},
  title   = {Climate change policy: What do the models tell us?},
  journal = {Journal of Economic Literature},
  volume  = {51},
  number  = {3},
  pages   = {860--872},
  doi     = {10.1257/jel.51.3.860},
  url     = {https://www.aeaweb.org/articles?id=10.1257/jel.51.3.860}
}

@article{Enkvist2007CostCurve,
  author  = {Enkvist, Per-Anders and Naucl{\'e}r, Tomas and Rosander, Jonas},
  year    = {2007},
  title   = {A cost curve for greenhouse gas reduction},
  journal = {The McKinsey Quarterly},
  number  = {1},
  pages   = {35--45}
}

@article{Leclere2020BendingCurve,
  author  = {Leclere, David and others},
  year    = {2020},
  title   = {Bending the curve of terrestrial biodiversity needs an integrated strategy},
  journal = {Nature},
  volume  = {585},
  number  = {7826},
  pages   = {551--556}
}

@article{Hof2017NDCCosts,
  author  = {Hof, Andries F. and den Elzen, Michel G. J. and Admiraal, Andries and Roelfsema, Mark and Gernaat, David E. H. J. and van Vuuren, Detlef P.},
  year    = {2017},
  title   = {Global and regional abatement costs of nationally determined contributions (NDCs) and of enhanced action to levels well below 2C and 1.5C},
  journal = {Environmental Science \& Policy},
  volume  = {71},
  pages   = {30--40},
  doi     = {10.1016/j.envsci.2017.02.008}
}

@misc{UN2022KunmingMontreal,
  author = {{United Nations}},
  year   = {2022},
  title  = {Kunming--Montreal global biodiversity framework},
  url    = {https://www.cbd.int/meetings/2022}
}

@misc{UKGov2021EnvironmentAct,
  author = {{UK Government}},
  year   = {2021},
  title  = {Environment Act 2021},
  url    = {https://www.legislation.gov.uk/ukpga/2021/30/contents/enacted}
}

@book{HMTreasury2022GreenBook,
  author    = {{HM Treasury}},
  year      = {2022},
  title     = {The green book: Central government guidance on appraisal and evaluation},
  publisher = {HM Treasury},
  address   = {London}
}

@misc{IPBES2022ValuesNature,
  author      = {{IPBES}},
  year        = {2022},
  title       = {Methodological assessment report on the diverse values and valuation of nature},
  editor      = {Balvanera, Patricia and Pascual, Unai and Christie, Michael and Baptiste, Brigitte and Gonz{\'a}lez-Jim{\'e}nez, Diego},
  institution = {IPBES Secretariat},
  address     = {Bonn},
  doi         = {10.5281/zenodo.6522522}
}

@article{Jochum2020BEFExperiments,
  author  = {Jochum, Margarete and Fischer, Markus and Isbell, Forest and Roscher, Christian and others},
  year    = {2020},
  title   = {The results of biodiversity--ecosystem functioning experiments are realistic},
  journal = {Nature Ecology and Evolution},
  volume  = {4},
  number  = {11},
  pages   = {1485--1494},
  doi     = {10.1038/s41559-020-1280-9}
}

@article{Horisch2024BiodiversityWTP,
  author  = {H{\"o}risch, Jacob and Petersen, Lars and Jacobs, Klaus},
  year    = {2024},
  title   = {The impact of biodiversity information on willingness to pay},
  journal = {Journal of Industrial Ecology},
  volume  = {28},
  number  = {6},
  pages   = {1641--1656},
  doi     = {10.1111/jiec.13552}
}

@article{Weitzman2009CatastrophicClimate,
  author  = {Weitzman, Martin L.},
  year    = {2009},
  title   = {On modeling and interpreting the economics of catastrophic climate change},
  journal = {Review of Economics and Statistics},
  volume  = {91},
  number  = {1},
  pages   = {1--19},
  doi     = {10.1162/rest.91.1.1}
}

@misc{DEFRA2023ENCA,
  author = {{Department for Environment, Food and Rural Affairs}},
  year   = {2023},
  title  = {Enabling a natural capital approach (ENCA): Guidance},
  url    = {https://www.gov.uk/government/publications/enabling-a-natural-capital-approach-enca-guidance}
}

@article{Aldy2021SCC,
  author  = {Aldy, Joseph E. and Kotchen, Matthew J. and Stavins, Robert N. and Stock, James H.},
  year    = {2021},
  title   = {Keep climate policy focused on the social cost of carbon},
  journal = {Science},
  volume  = {373},
  number  = {6557},
  pages   = {850--852},
  doi     = {10.1126/science.abi7813}
}

@article{Drupp2018Discounting,
  author  = {Drupp, Moritz A. and Freeman, Matthew C. and Groom, Ben and Nesje, Fredrik},
  year    = {2018},
  title   = {Discounting disentangled: An expert survey on the components of the long-term social discount rate},
  journal = {American Economic Journal: Economic Policy},
  volume  = {10},
  number  = {4},
  pages   = {109--134}
}

@article{Dietz2021TippingPoints,
  author  = {Dietz, Simon and Rising, James and Stoerk, Thomas and Wagner, Gernot},
  year    = {2021},
  title   = {Economic impacts of tipping points in the climate system},
  journal = {Proceedings of the National Academy of Sciences},
  volume  = {118},
  number  = {34},
  doi     = {10.1073/pnas.2103081118}
}

@article{DruppHansel2021RelativePrices,
  author  = {Drupp, Moritz A. and H{\"a}nsel, Martin C.},
  year    = {2021},
  title   = {Relative prices and climate policy},
  journal = {American Economic Journal: Economic Policy},
  volume  = {13},
  number  = {1},
  pages   = {168--201},
  doi     = {10.1257/pol.20180760}
}

@article{ManningAndo2022BatServices,
  author  = {Manning, David T. and Ando, Amy},
  year    = {2022},
  title   = {Ecosystem services and land rental markets: Producer costs of bat population crashes},
  journal = {Journal of the Association of Environmental and Resource Economists},
  volume  = {9},
  number  = {6},
  pages   = {1235--1277}
}

@article{duran_practical_2020,
	title = {A practical approach to measuring the biodiversity impacts of land conversion},
	volume = {11},
	issn = {2041-210X, 2041-210X},
	doi = {10.1111/2041-210X.13427},
	language = {en},
	number = {8},
	urldate = {2024-07-30},
	journal = {Methods in Ecology and Evolution},
	author = {Durán, América P. and Green, Jonathan M. H. and West, Christopher D. and Visconti, Piero and Burgess, Neil D. and Virah‐Sawmy, Malika and Balmford, Andrew},
	year = {2020},
	pages = {910--921},
}

@article{baumgartner_measuring_2006,
	title = {Measuring the {Diversity} of {What}? {And} for {What} {Purpose}? {A} {Conceptual} {Comparison} of {Ecological} and {Economic} {Biodiversity} {Indices}},
	issn = {1556-5068},
	shorttitle = {Measuring the {Diversity} of {What}?},
	doi = {10.2139/ssrn.894782},
	language = {en},
	urldate = {2024-07-30},
	journal = {SSRN},
	author = {Baumgärtner, Stefan},
	month = mar,
	year = {2006},
}

@article{Frank2024PestControl,
  author  = {Frank, Eyal},
  year    = {2024},
  title   = {The economic impacts of ecosystem disruptions},
  journal = {Science},
  doi     = {10.1126/science.adg0344}
}

\section*{Acknowledgements}

We thank Matthew Bardrick, Colin Smith, Alastair Johnson and Max Heaver from the UK Department of Environment, Food and Rural Affairs (DEFRA) for invaluable guidance in building the concepts in this paper. All remaining errors are our own.

\section*{Funding}

BG, BB and SM would like to thank Dragon Capital for funding the Dragon Capital Chair in Biodiversity Economics. BG is also funded by the UKRI/NERC RENEW project (NE/W004941/1) and the BIOPATH research programme funded by the Swedish Foundation for Strategic Environmental Research MISTRA (F 2022/1448). BG, BB, SM and FV also acknowledge generous financial support from the NERC/UKRI funded BIOADD grant [NE/X002292/1]. BG and FV acknowledge financial support from the Grantham Research Institute on Climate Change and the Environment, at the London School of Economics and the ESRC Centre for Climate Change Economics and Policy (CCCEP) via ESRC grant ref: ES/R009708/1. AB, BD, BG, IJB, MM, SM and DR would like to thank the DEFRA funded project entitled "Biodiversity Target and Cost Approach Scoping Project", OCID: 2025/S 000-070653. AB, BD, IJB, MM thank the UKRI funded LUNZ hub project: BB/Y008723/1. BG, IJB, EJMG, ABu, TA, ND, SzE and MM would like to acknowledge the NERC/UKRI-funded AGILE programme [NE/W004976/1]. SzE was additionally supported by EU Horizon 2020 project SUPERB (Systemic Solutions for Upscaling of Urgent Ecosystem Restoration for Forest Related Biodiversity) [GA 101036849], IJB, MM and AB by BBSRC NetZeroPlus [BB/V011588/1] and EJMG and ND by the Leverhulme Centre for Nature Recovery. ND thanks the NERC DTP. For the purpose of open access, the authors have applied a `Creative Commons Attribution (CC BY) license to any Author Accepted Manuscript version arising from this submission.

\section*{Author contributions}

\noindent \textbf{Initial ideas:} BG IJB JL SzE EJMG RH PS RS (Members of the UK Treasury Biodiversity Working Group) BD FV

\noindent \textbf{Conceptualization:} BG IJB JL SzE EJMG RH PS RS (Members of the UK Treasury Biodiversity Working Group) FV

\noindent \textbf{Methodology:} BG IJB BD SzE EJMG AB HM-P MM DR FV

\noindent \textbf{Investigation:} BG IJB JL SzE EJMG TA BB AB ABu BD ND RH HM-P MM SM HN DR RS PS FV

\noindent \textbf{Visualization:} BG AB HM-P MM FV

\noindent \textbf{Funding acquisition:} AB BG EJMG IJB 

\noindent \textbf{Writing -- initial draft:} BG IJB SzE EJMG AB HM-P MM DR

\noindent \textbf{Writing -- review \& editing:} BG IJB JL SzE EJMG BB AB BD RH HM-P MM SM DR RS PS FV

\section*{Competing interests}

The authors declare that they have no competing interests.

\section*{Data and materials availability}

Raw data and code for all figures and maps will be publicly available upon publication.

\section*{Supplementary information}

See attached PDF.

\end{document}